\documentclass[twocolumn,           % Format : preprint, twocolumn
               showpacs,            % Pacs : showpacs, noshowpacs
               preprintnumbers,     % Preprint: preprintnumbers,
               aps,                 % Society: ...
               prl,                 % Journal Style : pra, prb, prc, prd, pre,
               a4paper,             % Size : a4paper, ...
               superscriptaddress,  % Affiliation (Title) : groupedaddress,
               nofootinbib,         % Footnote: footinbib, nofootinbib
               tightenlines,        % Remove additional spaces in a line
               floats,floatfix      % Floating Pictures and tables
               ]{revtex4}

\usepackage{graphicx}% Include figure files
\usepackage{dcolumn}% Align table columns on decimal point
\usepackage{bm}% bold math
\usepackage{latexsym}
\usepackage{amsmath,amssymb}        % amssymb includes amsfonts
\usepackage[draft=false]{hyperref}

\usepackage[latin1]{inputenc}
\usepackage{latexsym}
\usepackage{amsmath}
\usepackage{epsfig}
\usepackage{ifpdf}

\begin{document}
\title{A brief Review of the Scalar Field Dark Matter model}
\author{Juan Maga\~na\footnote{Part of the Instituto Avanzado de Cosmolog\'ia (IAC) collaboration http://www.iac.edu.mx/}}
\email{jmagana@astroscu.unam.mx}
\affiliation{Instituto de Astronom\'{\i}a,
              Universidad Nacional Aut\'onoma de M\'{e}xico,
              AP 70-264, 04510 M\'exico D.F., M\'exico}

\author{Tonatiuh Matos$^{*}$}
\email{tmatos@fis.cinvestav.mx}
\affiliation{Departamento de F\'isica, Centro de Investigaci\'on y de
  Estudios Avanzados del IPN, A.P. 14-740, 07000 M\'exico D.F.,
  M\'exico.}

\author{Victor H. Robles$^{*}$}
\email{vrobles@fis.cinvestav.mx}
\affiliation{Departamento de F\'isica, Centro de Investigaci\'on y de
  Estudios Avanzados del IPN, A.P. 14-740, 07000 M\'exico D.F.,
  M\'exico.}

\author{Abril Su\'arez$^{*}$}
\email{asuarez@fis.cinvestav.mx}
\affiliation{Departamento de F\'isica, Centro de Investigaci\'on y de
  Estudios Avanzados del IPN, A.P. 14-740, 07000 M\'exico D.F.,
  M\'exico.}
\begin{abstract}
%%%%%%  Hydro %%%%%%%%%%
In the last time the cold dark matter (CDM) model has suggested more and more that it is not able to
describe all the properties of nearby galaxies that can be observed in great detail as well as that it has some problems 
in the mechanism by which matter is more rapidly gathered into large-scale structure such as galaxies and clusters of galaxies. 
In this work we revisit an alternative model, the scalar field dark matter (SFDM) model, 
which proposes that the galactic haloes form by condensation of a scalar field (SF) very early in the 
Universe,  i.e., in this model the haloes of galaxies are astronomical Bose-Einstein Condensate drops of SF. 
%forming Bose-Einstein Condensates (BEC) drops, i.e., in this model the haloes of galaxies are huge drops of SF. 
On the other hand, large-scale structures like clusters or superclusters of galaxies form similar to the $\Lambda$CDM  model, 
by hierarchy, thus all the predictions of the $\Lambda$CDM  model at cosmological scales are 
reproduced by SFDM. This model predicts that all galaxy haloes must be very similar and exist 
for higher redshifts than in the $\Lambda$CDM  model. 
In the first part of this review we revisit the cosmological evolution 
of SFDM model with a scalar potential  $m^2\Phi^2/2+\lambda\Phi^4/4$ with two different frameworks: the field and fluid approach.
We derive the evolution equations of the SF in the linear regime of perturbations as well. 
The scalar fluctuations have an oscillating growing mode and therefore, this kind of dark matter could lead to
the early formation of gravitational structures in the Universe.
We also revisit how BEC dark matter haloes exhibit a natural cut of the mass power spectrum.
%%%%%%%%%%%  Flat %%%%%%%%%%%%%
In the last part, we study the core central density profiles of BEC dark matter haloes and 
fit high-resolution rotation curves, we show a sample of some low surface
brightness galaxies. The mean value of the logarithmic inner density slopes is 
$\alpha = - $0.27 $\pm$ 0.18. Using a model independent new definition of the core in the BEC density profile, 
we show that the recent observation of the constant dark matter central surface density can be reproduced. 
We conclude that in light of the difficulties that the $\Lambda$CDM model is currently
facing the SFDM model can be a worthy alternative to keep exploring further.
\end{abstract}

\date{\today}


\maketitle

%\section{Hydro}
\section{Introduction}

Nowadays the most accepted model in cosmology which explains the evolution of the Universe is known as 
$\Lambda$CDM. In this model 4 per cent of the total content of the Universe is baryonic matter, 
22 per cent is non-baryonic dark matter (DM) and the rest is in some form of cosmological constant.
$\Lambda$CDM has achieved several observations with outstanding success, 
like for example the fact that the cosmic microwave background radiation can be explained 
in great detail and that it provides a framework within one can understand the large-scale 
isotropy of the Universe and important characteristics on the origin, nature and evolution of 
density fluctuations which are believed to give rise to galaxies and other cosmic structures. 
There remain, however, certain conflicts at galactic scales, like the cusp profile of central densities in galactic halos, 
the overpopulation of substructures predicted by $N$-body numerical simulations which are 
an order of magnitude larger than what has been observed, etc
(see for example \cite{clowe}, \cite{klypin}, \cite{moore}, and \cite{penny}).

There are other important issues of the $\Lambda$CDM model which require further considerations. 
Observations point out to a better understanding of the theory beginning with the Local Void, 
which contains far fewer galaxies than the expected. Another problem arises for the so called pure disk galaxies, 
which do not appear in numerical simulations of structure formation in the standard model.
These problems would be solved, if the structure grew faster than it does in the standard paradigm \cite{peebles}.
%because the relatively slow assembly allows stars to accumulate in thick stellar bulges 

On the other hand, \citet{lee_komatzu} also found that the collision velocity of 3000km/s at R200 for the Bullet Cluster 
is very unlikely and might no be explained by the standard model.
A final example of inconsistencies can be seen in a paper by 
\citet{Shaun}, who found anomalies in the mass power spectrum obtained by SDSS and the one obtained with the CDM model.
With these and other results it seems necessary to change the $\Lambda$CDM paradigm to try and explain the formation of structure in the Universe.

The incorporation of a new kind of DM, different from the one proposed by the $\Lambda$CDM 
model into the big bang theory holds out the possibility of resolving some of these issues.
Several authors have proposed interesting alternatives in where they
try to solve the latter explained difficulties that the CDM scenario seem not solve.
Some of these models consider warm dark matter (\cite{colin}, \cite{villa}), self interacting dark matter (\cite{loeb},\cite{spergel}),
and other exotic scenarios as braneworld models \citep{miguel}.
In fact, there are models that do not include DM but instead modify the Newtonian force law \cite{milgrom,sanders}. 
Other alternative scenario that has received much attention in the last years is the scalar field dark matter (SFDM) model. 
The main idea is simple (see \citet{guzmanII}), the nature of the DM is completely determined by a fundamental scalar field $\Phi$. 
The SFDM model proposes that galactic haloes form by Bose-Einstein condensation of a scalar field (SF) whose boson
has an ultra-light mass of the order of $m\sim 10^{-22}$eV. From this mass it follows that the critical temperature of condensation 
$T_c\sim1/m^{5/3}\sim$TeV is very high, therefore, they form Bose-Einstein Condensates (BEC) drops very early in the Universe. 
In addition, the Compton length $\lambda_{c}=2 \pi \hbar/m$ associated to this boson is about $\sim$~kpc that corresponds to 
the dark halo-size of typical galaxies in the Universe. Thus, it has been proposed that these drops
are the haloes of galaxies (see \citet{further}), $i.e.$, that haloes are huge drops of SF. 
On the other hand, big structures form like in the $\Lambda$CDM  model, by hierarchy \citep{further,suarez}, thus, all 
successful predictions of the standard  model at large scales are well reproduced by SFDM. 
In other words, in the SFDM model the haloes of galaxies do not form hierarchically, 
they are formed at the same time and in the same way when the Universe reaches the critical temperature
of condensation of the SF. From this it follows that all galaxies must be
very similar because they formed in the same manner and at the same moment \citep{leelim}. 
Therefore, from this paradigm we have to expect that 
there exist well formed galaxy haloes at higher redshifts than in the $\Lambda$CDM  model.  
In this model the scalar particles with that ultra light mass are such that their wave properties 
avoid the cusp problem and reduce the high number of small satellites by the quantum uncertainty principle \citep{hu, lundgren}
which is another problem that is still present in the $\Lambda$CDM model \citep{harko_core, klypin, victor, }. Summarizing, it is remarkable that
with only one free parameter, the ultra-light scalar field mass ($m \sim 10^{-22}$eV), the SFDM model fits:

\begin{enumerate}
\item The evolution of the cosmological densities \citep{phi2}.

\item The acoustic peaks of the cosmic microwave background \citep{ivan}. 

\item The scalar field has a natural cut off, thus the substructures in clusters of galaxies is avoided naturally. 
With a scalar field mass of $m_\phi\sim10^{-22}$eV the amount of substructure is compatible with the observed one \citep{further,hu,suarez}.

\item We expect that SFDM forms galaxies earlier than the CDM model, because they form BECs at a 
critical temperature $T_c >> $TeV. So if SFDM is right, we have to see big galaxies at high redshifts with similar features 
\citep{phi2,suarez}.  

\item The rotation curves of big galaxies and LSB galaxies \citep{Lesgourgues:2002hk,Arbey:2003sj,harkomnras11,argelia, victor}.

\item With this mass, the critical mass of collapse for a real scalar field is just $10^{12}\,M_{\odot}$, i.e., the one 
observed in galaxies haloes \citep{alcubierre}.

\item The observed properties of dwarf galaxies, i.e., the minimum length scale, the minimum
mass scale, and their independence from the brightness \citep{leelim}.

\item And recently it has been demonstrated that if the scalar mass is $m\sim10^{-22}$eV,
the SFDM haloes would have cores large enough to explain the longevity of the cold clump in Ursa Minor
and the wide distribution of globular clusters in Fornax
(see \citet{lora}).
\end{enumerate}

The idea was first considered by \citet{sin,ji_sin} and independently by
\citet{L9,guzmanI,guzmanII,lee_koh} suggested bosonic dark matter as a model for galactic halos \citep{dehnen}. 
In the BEC model, DM haloes can be described, in the non-relativistic regime, as Newtonian gravitational condensates
made up of ultra light bosons condensated in a single macroscopic wave function. 
Thus, BECs haloes can be described as a coherent scalar field $\Phi$.

Several authors have introduced a dynamic scalar field with a certain potential $V(\Phi)$ as a candidate to dark matter, 
although there is not yet an agreement for the correct form of the potential of the field. 
One interesting work pointing this way was done by \cite{mat_ure2000} and independently by \cite{sahni} 
where they used a potential of the form $V(\Phi)=V_{0}\left[\cosh\left( \xi \Phi \right)-1\right]$
to explain the core density problem for disc galaxy halos in the $\Lambda$CDM model 
(see also \cite{Wetterich}, \cite{Ratra} and \cite{Copeland}). 
\citet{guzmanII} presented a model for the DM in spiral galaxies, in which they supposed that DM
is an arbitrary SF endowed with a scalar potential. Other scalar potential widely used to describe DM is
$V(\Phi)=m^{2}\Phi^{2}/2$ \citep{turner,phi2}. This potential is very interesting because it can mimic
the cosmological evolution of the Universe predicted by the $\Lambda$CDM model.
If we consider a SF self-interaction, we need to add a quartic term to the SF potential \citep{Arbey:2001qi, Arbey:2005zc,Arbey:2006it, Arbey:2008um,fabio},
in this case the equation of state of the SF is that of a polytope of index n=1 (see \citep{suarez,harkomnras11,harko_cosmo,Arbey:2001jj, Lesgourgues:2002hk}). 

Different issues of the cosmological behavior of the SFDM/BEC model have been studied
(see for example \citep{further,Arbey:2001qi,hu,kainb,luis_bose,lundgren,marsh,ivan, ivan2, suarez, harko_cosmo, harkomnras11, phi2, Arbey:2005zc,matos_open,chavanisIII,velter,woo}).
\citet{hu} proposed fuzzy dark matter that is composed of ultra light scalar particles who are initially in the form of a BEC.
In their work \citet{woo} used a bosonic dark matter model to explain the structure formation via high-resolution simulations, 
\citet{urenaAIP, luis_bose} reviewed the key properties that may arise from the bosonic nature of SFDM models.
Recently \citet{harko_cosmo, harkomnras11} developed a further analysis of the cosmological dynamics of SFDM/BEC 
as well as of the evolution of their fluctuations (see also \cite{chavanisIII}). In the same direction, \citet{suarez} studied,
the growth of scalar fluctuations and large-scale structure formation with the fluid approach of the SFDM/BEC model. 

On the other hand, SFDM/BEC model has provided to be a
good candidate for DM haloes of galaxies in the Universe
because it can explain many aspects where the standard model fails 
(\citep{argelia,boharko,lee2,leelim,leelimchoi,lora,harko_core,victor}).
In addition, many numerical simulations have been performed to study the gravitational collapse of SFDM/BEC model
\citep{argelia2,argelia3,colpi,fcoluis, colapsob, apjpaco, gleiser}.
\citet{chavanisI,chavanisII} found approximate analytical expression
and numerical solutions of the mass-radius relation of SFDM/BEC haloes. 
Recently, \citet{rindler} give constraints on the boson mass to form and mantain more than one vortex in SFDM/BEC haloes.
These constraints are in agreement with the ultra light mass founded in previous works (see also \citep{kaina,zinner}).
Lately, \citet{lora} performed N-body simulations to study the dynamics observed in Ursa Minor dwarf galaxy. 
They model the dark matter halo of Ursa Minor as a SFDM/BEC halo to establish constraints for the boson mass.
Moreover, they introduce a dynamical friction analysis with the SFDM/BEC model 
to study the wide distribution of globular clusters in Fornax. An overall good agreement is found
for the ultra light mass $\sim 10^{-22}$eV of bosonic dark matter.

The main objective of this review is to introduce the framework of the 
SFDM/BEC model that assumes that DM is a scalar field that 
involves a self-interacting potential of the form $V(\Phi)=m^2\Phi^2/2+\lambda\Phi^4/4$, 
where $m\sim 10^{-22}$eV is the mass of the scalar field, \cite{lee_koh}, \cite{further} and \cite{hu}. 

We solve the Friedmann equations for the SFDM/BEC model to show that it behaves
just like the CDM model. Also, we show that the SFDM/BEC predicts galaxy formation earlier than the CDM model, because they form BEC at a critical 
temperature $T_c >>$ TeV. So, if SFDM is right, this would imply that we have to see big galaxies at high redshifts. 
In order to do this, we study the density fluctuations of the scalar field from a hydrodynamical point of view, 
this will give us some information about the energy density of dark matter halos necessary to obtain the observational results 
of large-scale structure. Here we will give some tools that might be necessary for the study of the early formation of structure.

In section \ref{cosmology} we study the cosmological evolution of the SF with a field and a fluid approach.
We obtain a numerical solution for the scalar density as well as analytical expressions for
the kinetic and potential energies of the SF. Also, we study the evolution for the density contrast
of scalar fluctuations in the linear regime. In section \ref{galacticSFDM}
we study the rotation curves and the cusp/core discrepancy with SFDM haloes to
compare with the predictions of CDM. Finally we give our conclusions in section \ref{sec:conclusions}.

\section{The Scalar Field Dark Matter Cosmology} \label{cosmology}
The framework of the standard cosmological model is of a homogeneous and isotropic Universe whose evolution 
is best described by Friedmann's equations that come from general relativity and whose main ingredients 
can be described by fluids whose characteristics are very similar to those we see in our Universe. 
Of course, the Universe is not exactly homogeneous and isotropic but this standard model does give us a 
framework within which we can study the evolution of the expansion rate of the Universe as well as
the evolution of small fluctuations that give rise to gravitational structures in the Universe as 
galaxies and clusters of galaxies.

In this section, we study the cosmological dynamics of the SFDM/BEC model and the evolution
of their scalar perturbations with two different frameworks: the field and fluid approach. 

\subsection{Background Universe: The Field Approach}
\label{sec:background}

We use the Friedmann-Lema\^itre-Robertson-Walker (FLRW) metric with scale factor $a(t)$. Our background Universe is composed by
SFDM ($\Phi_{0}(t)$) endowed with a scalar potential $V\equiv V(\Phi_0)$, radiation ($z$), neutrinos ($\nu$), baryons ($b$), and a cosmological constant ($\Lambda$) as dark energy. 
We begin by recalling the basic background equations. From the energy-momentum tensor $\mathbf{T}$ for a scalar field, the scalar energy density $T_0^0$ and the scalar pressure $T_j^i$ are given by
\begin{equation}
T_0^0=-\rho_{\Phi_0}=-\left(\frac{1}{2}\dot{\Phi}_0^2+ V\right),
\label{rhophi0}
\end{equation}
\begin{equation}
T_j^i=P_{\Phi_0}=\left(\frac{1}{2}\dot{\Phi}_0^2-V\right)\delta_j^i,
\label{pphi0}
\end{equation}
where the dots stand for the derivative with respect to the cosmological time and $\delta^i_j$ is the Kronecker delta. Thus, the Equation of State (EoS) for the scalar field is $P_{\Phi_{0}}=\omega_{\Phi_{0}}\,\rho_{\Phi_0}$ with
\begin{equation}
\omega_{\Phi_{0}}= \frac{\frac{1}{2}\dot{\Phi}_{0}^{2}\,-\,V}{\frac{1}{2}\dot{\Phi}_{0}^{2}\,+\,V}.
\label{ec:w}
\end{equation}

The radiation fields, the baryonic component and the cosmological constant are represented by perfect fluids with baryotropic equation of state $P_{\gamma}=(\gamma-1)\rho_{\gamma}$, where $\gamma$ is a constant, $0\le \gamma \le 2$. For example, $\gamma_{z}=\gamma_{\nu}=4/3$ 
for radiation and neutrinos, $\gamma_{b}=1$ for baryons, and for a cosmological constant $\gamma_{\Lambda}=0$.

The Einstein-Klein-Gordon equations that describe this Universe are (in units $c=\hbar=1$).
\begin{subequations}
\begin{eqnarray}
\dot H&=&-\frac{\kappa^2}{2}\left(\dot{\Phi}_{0}^2+\frac{4}{3}\rho_z+\frac{4}{3}\rho_{\nu}+
\rho_{b}\right),\\
\ddot{\Phi}_{0} &+& 3\,H \dot{\Phi}_{0}+ V,_{\Phi_{0}}=0,\label{eq:nopertKG}\\
{\dot\rho_{z}}&+&4\,H \rho_{z}=0,\\
{\dot\rho_{\nu}}&+&4\,H \rho_{\nu}=0,\\
{\dot\rho_{b}}&+& 3\,H \rho_{b}=0, 
\end{eqnarray}\label{eq:back}
\end{subequations}
with the Friedmann constraint
\begin{equation}
H^2=\frac{\kappa^2}{3}\left(\rho_{\Phi_{0}}+\rho_{z}+
\rho_{\nu}+\rho_{b} + \rho_{\Lambda} \right),
\label{eq:FC}
\end{equation}
\noindent
being $\kappa^{2} \equiv 8\pi G$, $H \equiv \dot{a}/a$ the Hubble parameter and the commas stand for the derivative with respect to scalar field. Notice that background scalar quantities at zero order have the subscript $0$.
%\subsection{The rescaled equations for the background Universe}
In order to solve the system of equations (\ref{eq:back}), we define the following dimensionless variables
\begin{eqnarray}
x&\equiv& \frac{\kappa}{\sqrt{6}}\frac{\dot{\Phi}_{0}}{H},\quad
u \equiv \frac{\kappa}{\sqrt{3}}\frac{\sqrt{V}}{H},\nonumber\\
z&\equiv& \frac{\kappa}{\sqrt{3}}\frac{\sqrt{\rho_{z}}}{H},\,\,\,
\nu \equiv \frac{\kappa}{\sqrt{3}}\frac{\sqrt{\rho_{\nu}}}{H},\nonumber\\
b&\equiv& \frac{\kappa}{\sqrt{3}}\frac{\sqrt{\rho_{b}}}{H},\,\,\,
l \equiv \frac{\kappa}{\sqrt{3}}\frac{\sqrt{\rho_{\Lambda}}}{H}.
\label{eq:varb}
\end{eqnarray}

Here we take the quadratic scalar potential $V=m^{2}\Phi_{0}^{2}/2$ with $m \sim1 \times10^{-22}$eV. Using these variables, the equations (\ref{eq:back}) for the evolution of the background Universe are transformed into
\begin{subequations}
\begin{eqnarray}
x'&=& -3\,x - s u+\frac{3}{2}\Pi\,x, \label{eq:dsb_x}\\
u'&=& s x +\frac{3}{2}\Pi\,u, \label{eq:dsb_u}\\
z'&=&\frac{3}{2}\left(\Pi-\frac{4}{3} \right)\,z, \label{eq:dsb_z}\\
\nu'&=&\frac{3}{2}\left(\Pi-\frac{4}{3} \right)\,\nu, \label{eq:dsb_nu}\\
b'&=&\frac{3}{2}\left(\Pi -1\right)\,b, \label{eq:dsb_b}\\
l'&=&\frac{3}{2}\Pi\,l, \label{eq:dsb_l}\\
s'&=&s_0 \,s^{-k}, \label{eq:dsb_s}
\end{eqnarray}\label{eq:dsb}
\end{subequations}
where the prime denotes a derivative with respect to the e-folding number $N=\ln a$, and $\Pi$ is defined as
\begin{equation}
-\frac{\dot H}{H^2}=\frac{3}{2}(2x^2+b^2+\frac{4}{3} z^2+\frac{4}{3} \nu^2) \equiv \frac{3}{2}\Pi. 
\end{equation}

We have introduced the variable $s\equiv C_{0}/H$, where $C_{0}$ is a constant, as a control parameter for the dynamics of $H$. 
In Eq. (\ref{eq:dsb_s}), $s_0$ is a constant and the exponent $k$ is $\le0$ (see \cite{mayra} for more details of the control parameter). 
With these variables, the density parameters $\Omega_{i}$ for each component $i$ can be written as
\begin{eqnarray}
\Omega_{\Phi_{0}}&=&x^2+u^2, \nonumber\\
\Omega_{z}&=&z^2,\nonumber\\
\Omega_{\nu}&=&{\nu}^2,\nonumber\\
\Omega_{b}&=&b^2,\nonumber\\
\Omega_{\Lambda}&=&l^2,\label{eq:dens}
\end{eqnarray}
subject to the Friedmann constraint
\begin{equation}\label{eq:fri}
x^2+u^2+z^2+\nu^{2}+b^2+l^2=1.
\end{equation}
In addition, we may write the EoS of the scalar field as
\begin{equation}
\omega_{\Phi_{0}}=\frac{x^{2}-u^{2}}{\Omega_{\Phi_{0}}}.
\label{eq:dlw}
\end{equation}
Since $\omega_{\Phi_{0}}$ is a function of time, if its temporal average tends to zero, this would imply that $\Phi^2$-dark matter can be able to mimic the EoS for CDM.

\begin{figure}
 \centering
 \scalebox{0.3}{\includegraphics{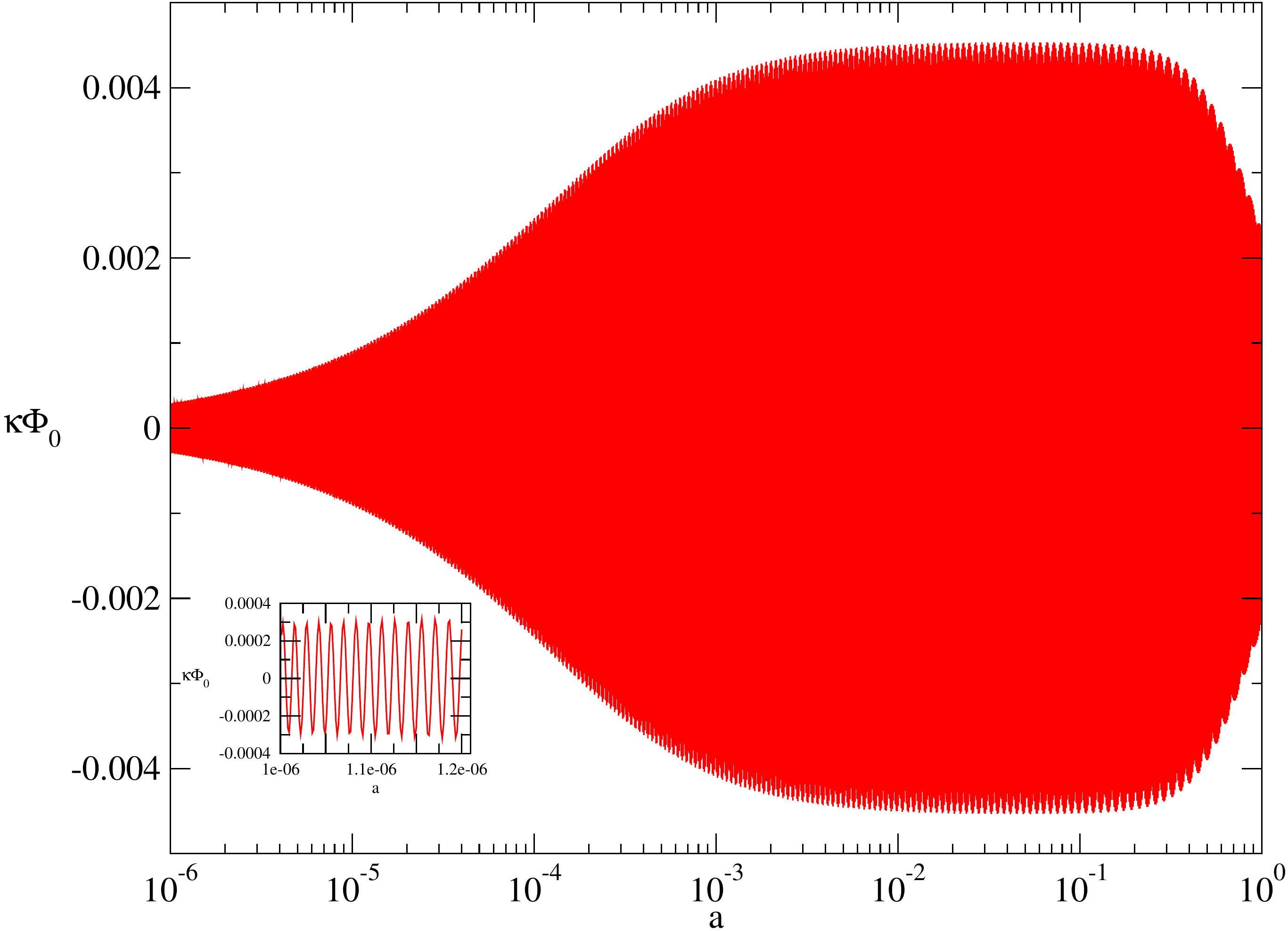}}
 \caption{Evolution of the scalar field $\Phi_{0}$ for the background Universe. The inset shows the fast oscillations of the scalar field.}
 \label{fig:phi}
\end{figure}

\begin{figure}
 \centering
 \scalebox{0.3}{\includegraphics{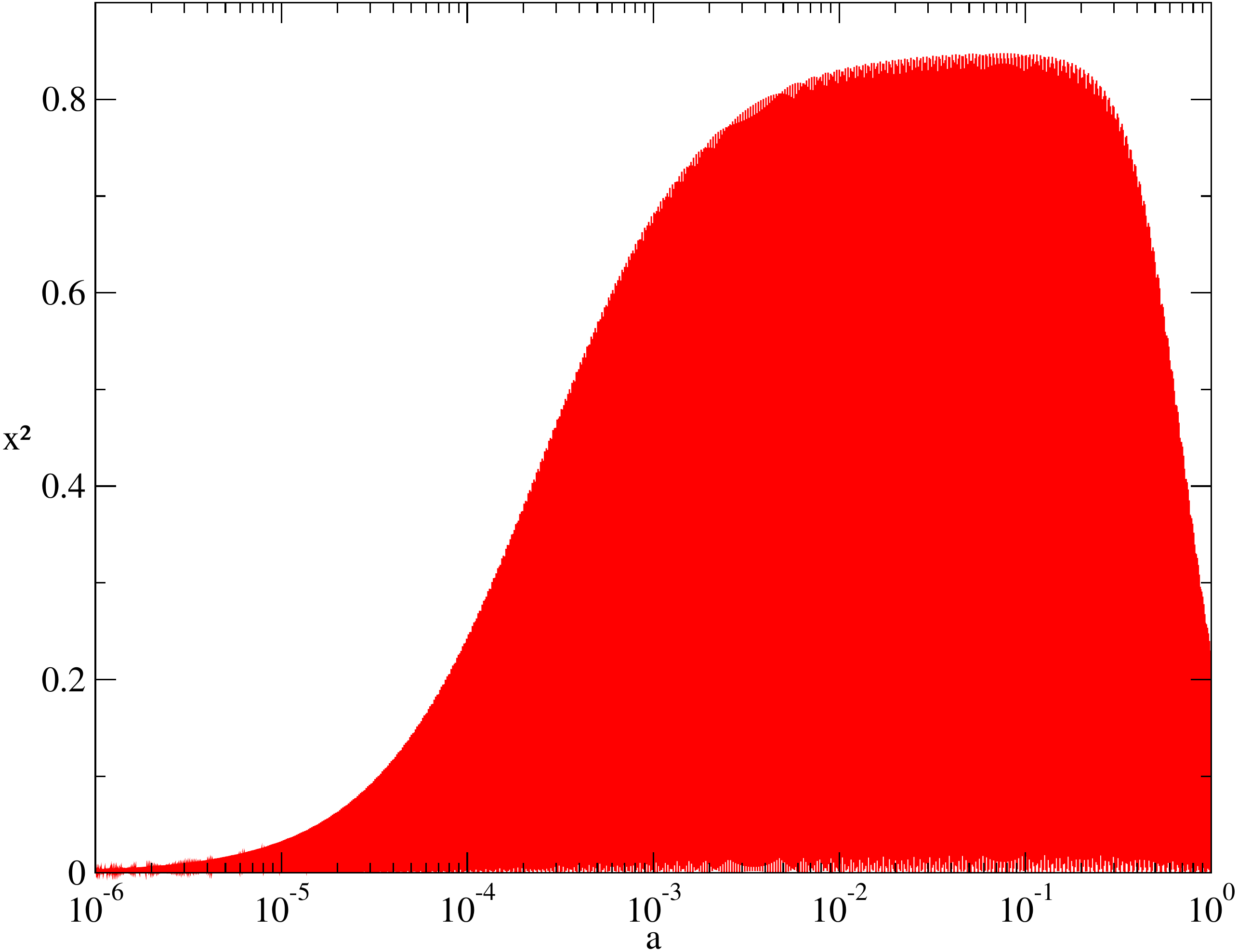}}
  \scalebox{0.3}{\includegraphics{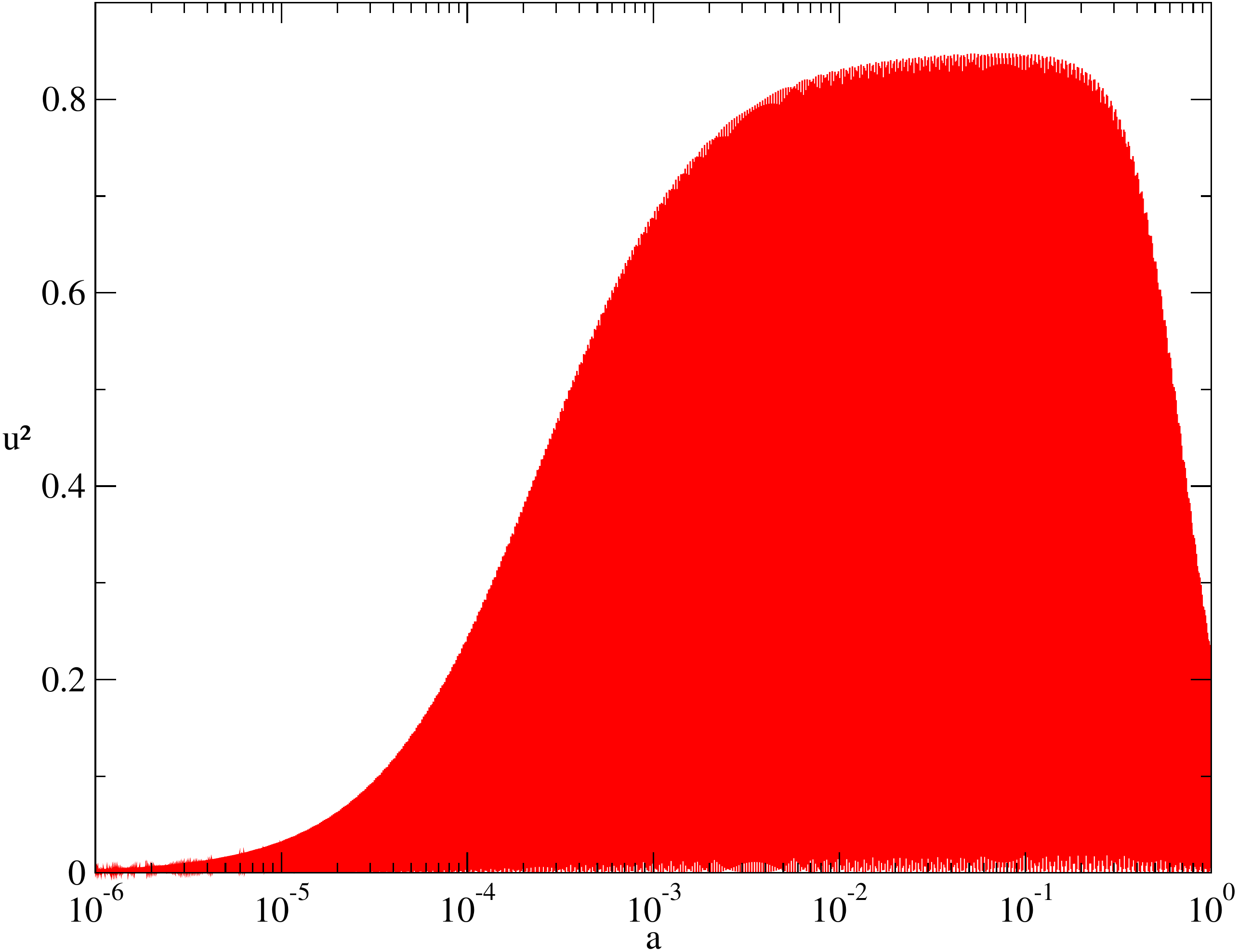}}
 \caption{Evolution of the kinetic (top panel) and potential (bottom panel) energy of the scalar field $\Phi_{0}$.}
 \label{fig:Ox}
\end{figure}

\begin{figure}[h]
 \centering
 \scalebox{0.3}{\includegraphics{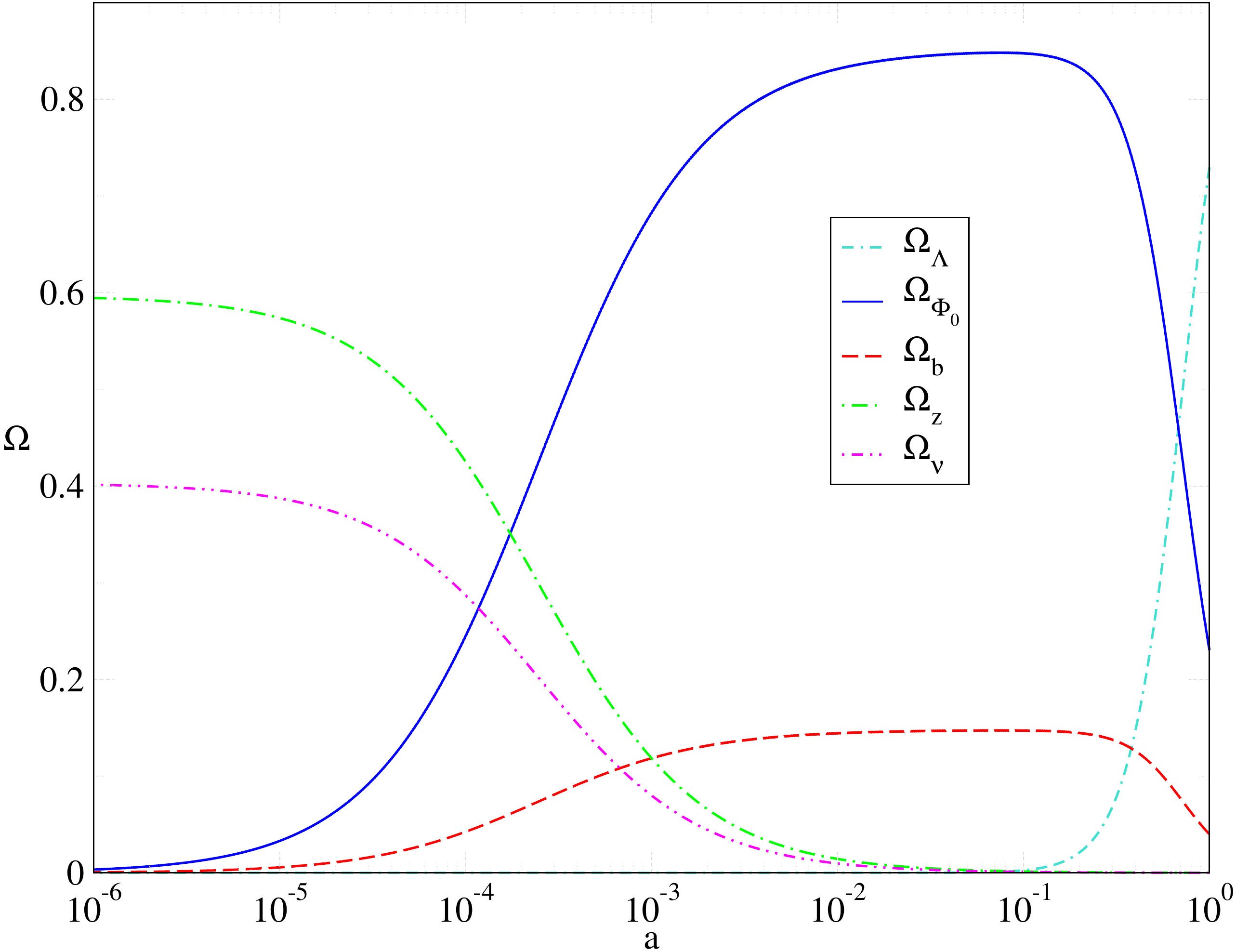}}
 \caption{Evolution of the density parameters $\Omega_{i}$ for the background Universe. SFDM model mimics the standard 
          $\Lambda$CDM behavior.}
 \label{fig:back}
\end{figure}

We solve the system of equations (\ref{eq:dsb}) for the background Universe numerically  with a four order Adams-Bashforth-Moulton (ABM) method. 
We take as initial conditions the best estimates from 5-years \cite{WMAP5} and 7-years WMAP \cite{WMAP7} values  
to $\Omega^{(0)}_{\Lambda}=0.73$, $\Omega^{(0)}_{DM}=0.22994$, $\Omega^{(0)}_b=0.04$, $\Omega^{(0)}_z=0.00004$, $\Omega^{(0)}_{\nu}=0.00002$ 
and the exponent $k=0$. In Fig.\ref{fig:phi} we see the evolution of the scalar field $\Phi_{0}$. 
This figure shows how the scalar field oscillates very stark about the minimum of the scalar potential $V$, 
on the bottom of this figure we show a zoom of these oscilations. In Fig.\ref{fig:Ox} we show the evolutions 
of the kinetic ($\dot\Phi^2/2$) and potential energy ($m^2\Phi^2/2$) of the scalar field, as expected, 
the SF oscilations are translated into very stark oscilations for the kinetic and the potential energies of the SF. 
However, observe the evolution of the dark matter density of the scalar field, 
that means $\rho_\Phi=(\dot\Phi^2+m^2\Phi^2)/2$ in Fig.\ref{fig:back}. 
Note the following crucial point, although the kinetic and potential energies show very stark oscillations, 
the sum of both energies does not oscillates at all, this sum is the density parameter $\Omega_{\Phi_{0}}$, 
which does not display any oscillation. The important point we have to take from now on into account is that the oscillations 
are not physical observables at all, they are a feature of the SF, 
what we observe in fact is the density of the SF which does not oscillates. 

\begin{figure}[h]
 \centering
 \scalebox{0.3}{\includegraphics{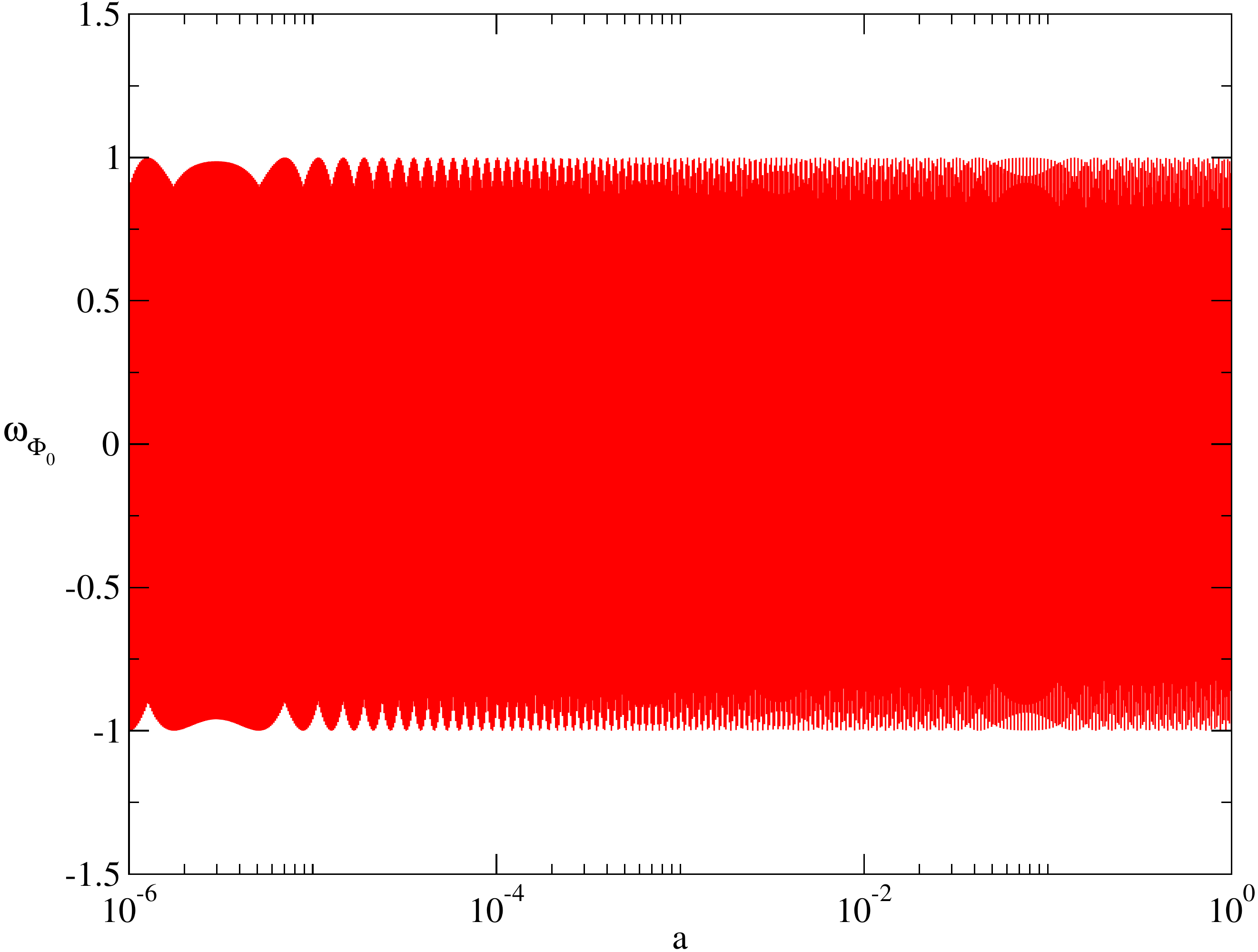}}
 \caption{Evolution of the SFDM equation of state for the background Universe.}
 \label{fig:w0}
\end{figure}

Fig.\ref{fig:back} shows the numerical evolution of the density parameters in our model\footnote{The scale factor is such that $a=1$ today, so that it relates to the redshift $z$ by $a=(1+z)^{-1}$.}. 
At early times, radiation dominates the evolution of the Universe. Later on, the Universe has an epoch where the energy density radiation
is equal to the dark matter density, at $z_{eq}$, then dark matter begins to dominate the evolution. 
The recombination era in SFDM model occurs at $z\sim1000$. At later times, the cosmological constant dominates 
the dynamics of the Universe at $z_{\Lambda} \sim 0.5$. The behavior is exactly the same as in the $\Lambda$CDM model.
Furthermore, in the recombination era of the SFDM model, the neutrinos made up $\sim12$ per cent
of the Universe completely in agreement with the measurements of WMAP.
Fig.\ref{fig:w0} shows the evolution of the EoS for the SF. Although the EoS varies with time (oscillations), 
the temporal average, $\left<\omega_{\Phi_0}\right>$, drops to zero. 
Therefore, SFDM is like a pressureless fluid and behaves as CDM at cosmological scales (\cite{turner}, \cite{further}, \cite{phi2}, \cite{matos_open}).
\subsection{Background Universe: The Fluid Approach}\label{fondo}

In this section we perform a transformation in order to solve the Friedmann equations analytically 
with the approximation $H<<m$. 

Here we take the scalar potential as $V=m^{2}\Phi^2/2\hbar^2+\lambda\Phi^4/4$, (we remain that we use units where $c=\hbar=1$ unless it is stipulated). Then for the ultra-light boson particle we have that $m\sim 10^{-22}$ eV.

Now we express the SF, $\Phi_0$, in terms of the new variables $S$ and $\hat\rho_0$, where $S$ is constant in the background and $\hat{\rho}_0$ will be the energy density of the fluid also in the background. So, our background field is proposed as
 \begin{equation}
  \Phi_0=(\psi_0\, e^{-\, imt/\hbar}+\psi^*_0\, e^{\, imt/\hbar})
 \end{equation}
where,
 \begin{equation}
  \psi_0(t)=\sqrt{\hat{\rho}_0(t)}\, e^{\, iS/\hbar}
 \end{equation}
and with this our SF in the background can be finally expressed as,
 \begin{equation}
  \Phi_0=2\sqrt{\hat{\rho}_0}\cos(S-mt/\hbar),
 \label{tri}
 \end{equation}
with this we obtain
 \begin{eqnarray}
  \dot{\Phi}_0^2&=&\hat{\rho}_0\left[\frac{\dot{\hat{\rho}}_0}{\hat{\rho}_0}\cos(S-mt/\hbar)\right.\nonumber\\
  &-&\left.2(\dot{S}-m/\hbar)\,\sin(S-mt/\hbar)\right]^2.
 \label{backSF}
 \end{eqnarray}

 To simplify, observe that the uncertanty relation implies that $m\Delta t\sim\hbar$, 
and for the background in the non-relativistic case 
the relation $\dot S/m\sim0$ is satisfied. Notice also that for the background we have that 
the density goes as $(\ln\hat{\rho}_0)\dot{}=-3H,$ but we also have that $H\sim 10^{-33}$ eV $<<m\sim 10^{-22}$ eV, so with these considerations at hand for the background, in (\ref{backSF}) we have
 \begin{equation}
  \dot{\Phi}_0^2=4\frac{m^2}{\hbar^2}\hat{\rho}_0\sin^2(S-mt/\hbar)
 \end{equation}
 
Finally, substituting this last equation and equation (\ref{tri}) into (\ref{rhophi0}) when taking $\lambda=0$, we obtain
  \begin{equation}
   \rho_{\Phi_0}=2\frac{m^2}{\hbar^2}\hat{\rho}_0[\sin^2(S-mt/\hbar)+\cos^2(S-mt/\hbar)]=2\frac{m^2}{\hbar^2}\hat{\rho}_0.
  \label{trigo}
  \end{equation}

Comparing this result with (\ref{eq:dens}) we have that the identity $\Omega_{\Phi_0}=2m^2\hat{\rho}_0/\hbar^2$ holds for the background, 
so comparing with (\ref{trigo}),
 \begin{equation}
  x=\sqrt{2\hat{\rho}_0}\frac{m}{\hbar}\sin(S-mt/\hbar)
 \label{eq:cinet}
 \end{equation}
 \begin{equation}
  u=\sqrt{2\hat{\rho}_0}\frac{m}{\hbar}\cos(S-mt/\hbar).
 \label{eq:potencial}
 \end{equation}

\begin{figure}
% \vspace{174pt}
 \scalebox{0.6}{\includegraphics{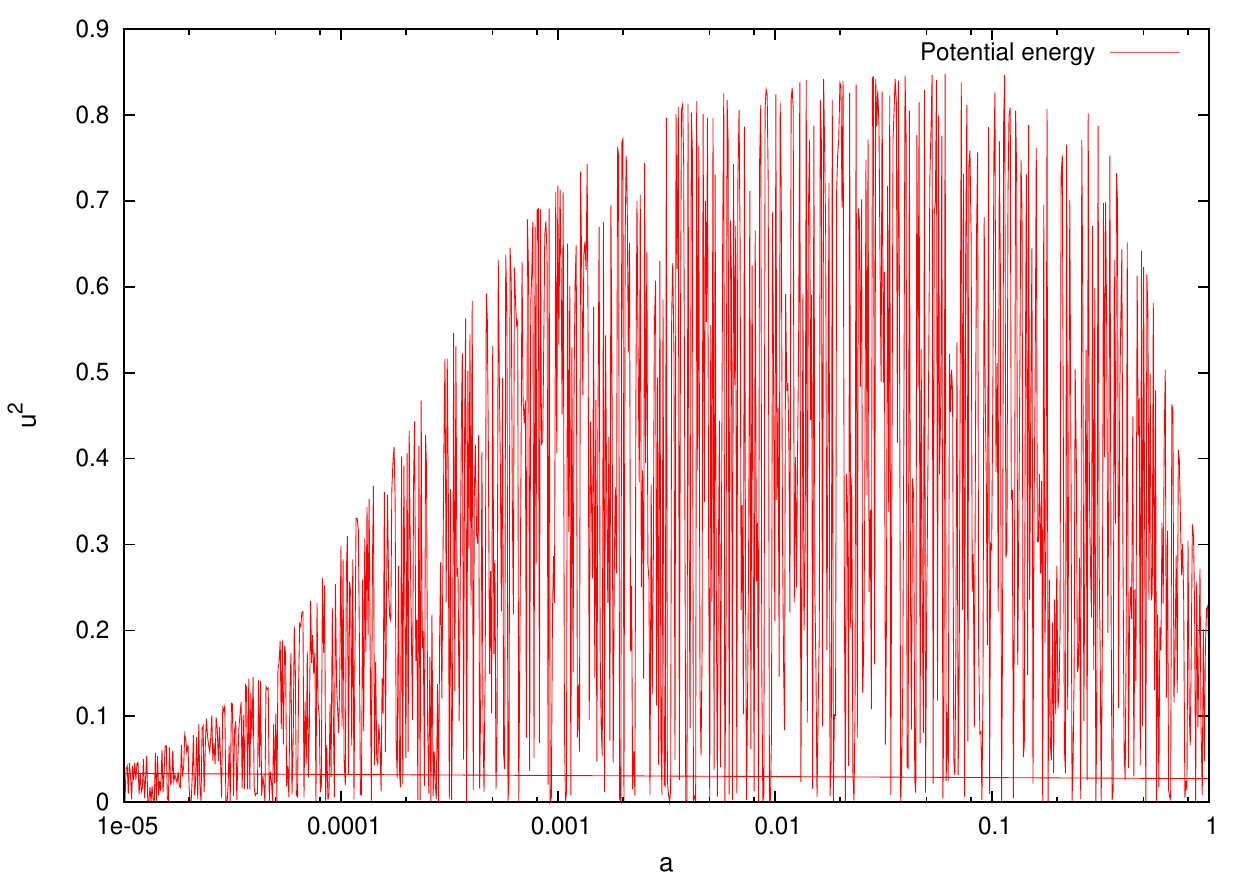}}
 \scalebox{0.6}{\includegraphics{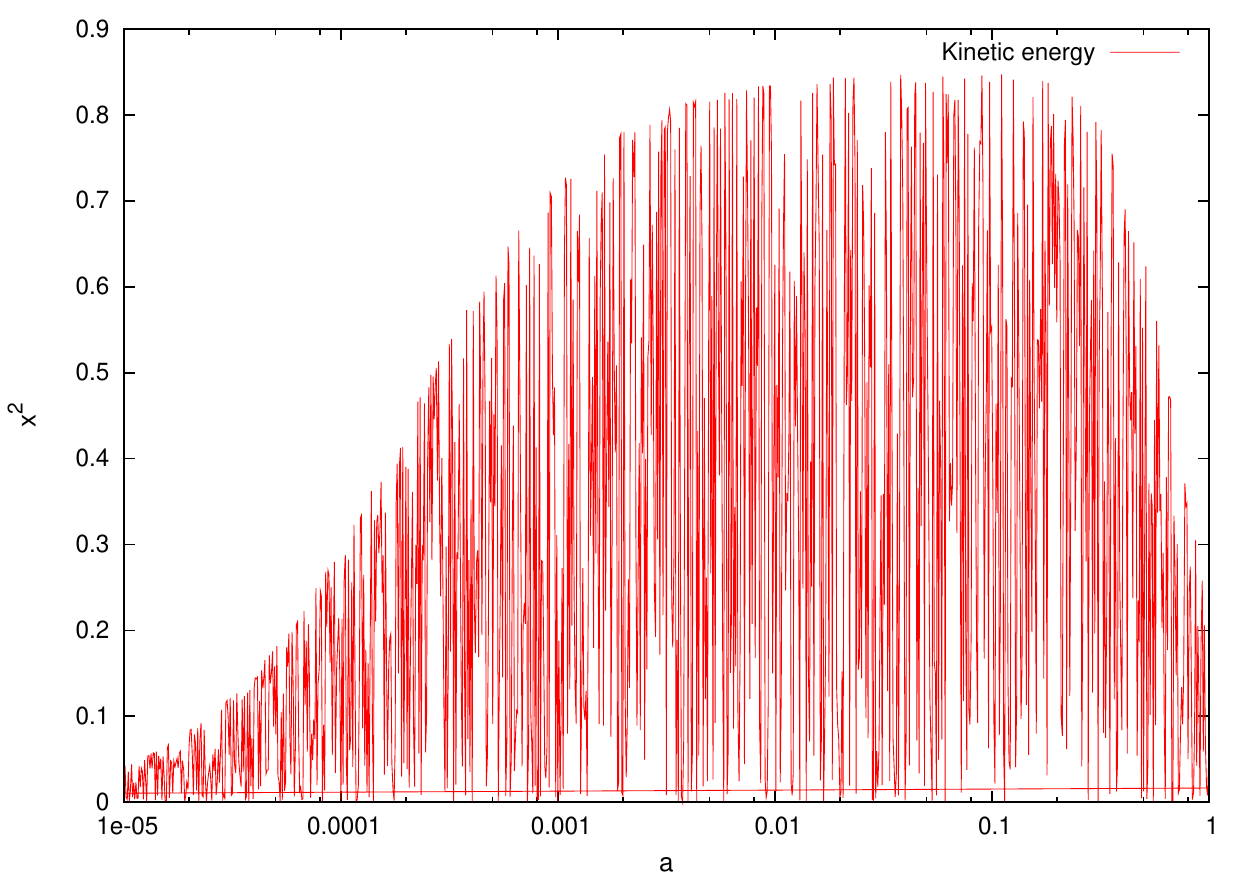}}
 \caption{Analytical evolution of the potential (top panel) and kinetic (bottom panel) energies of the scalar field dark matter.}
 \label{fig1}
 \end{figure}

In terms of the two analytic results, we show the evolution of the kinetic and potential energies (\ref{eq:cinet}) and (\ref{eq:potencial}) 
in Fig. \ref{fig1}, where for the evolution we used the e-folding number $N$ and the fact that $a\sim t^n\rightarrow t\sim\,{e}^{N/n}$. 

Observe the excelent accordance with the numerical results in \cite{phi2} for the kinetic and potential energies of the background respectively.
Therefore, we expect that if we sum the analytical kinetic and the potential energies we will obtain
the same behavior for the SFDM density shown in Fig. \ref{fig:back}.
In what follows we will use the analytic expressions (\ref{eq:cinet}) and (\ref{eq:potencial}) instead of the numerical ones.

\section{Scalar Field  Dark Matter Fluctuations}

Nowadays it is known that our Universe is not exactly isotropic and spatially homogeneous like the FLRW metric describes. 
There exist small deviations of this model that were the cosmological seeds to lead
the large-scale structure formation in the Universe. 
In this section we compute the growth of the SFDM overdensities $\delta\rho_{\Phi}$
at the stage when the density contrast $\delta\equiv\delta\rho_{\Phi}/\rho_{\Phi_{0}}$ is much smaller than unity. 
It is believed that the Universe was almost uniform after inflation, with very small density contrast. 
As the Universe expanded, the small over-densities grew until they began to collapse, 
leading to the formation of structure in the Universe. 
Here we consider small enough deviations in the FLRW model, so that they can be treated by linear perturbation theory.

\subsection{The Linear Regime of SFDM perturbations: The Field approach}\label{sec:linear}
In the SFDM paradigm, the dark haloes of the galaxies are
BECs composed of ultra light scalar particles with $m\sim10^{-22}$eV
then it would be about $10^{68}m$ GeV$^{-1}$ scalar particles to follow in a single galaxy
and the occupation numbers in galactic haloes are so big that the 
DM behaves as a coherent classical scalar field $\Phi$ that obeys the Klein-Gordon equation $(\square^2+m^2/\hbar^2)\Phi=0$, 
where $\Box$ is the D'Alambertian. Therefore, to study the
structure formation in the Universe we describe a non-interacting SFDM model such that: 
i) it can describe it more as a field than as particles and  
ii) we find a function that only depends on the three spatial coordinates and time.

First, we introduce the perturbed metric tensor in the FLRW background, we only consider scalar perturbations.
We then give the equation of energy-momentum conservation and the Einstein field equations for the perturbed metric.

By definition, a perturbation done in any quantity, is the difference between its value in some event in real space-time, and its 
corresponding value in the background. So, for example for the SF we have
 \begin{equation}
  \Phi=\Phi_0(t)+\delta\Phi(\mbox{\boldmath$x$},t),
 \label{pert}
 \end{equation}
where the background is only time dependent, while the perturbations also depend on the space coordinates. 
Similar cases apply for the metric;
 \begin{eqnarray}
  g_{00}&=&-a^2(1+2\phi),\nonumber\\
  g_{0i}&=&a^2B,_i,\nonumber\\
  g_{ij}&=&a^2[(1-2\psi)\delta_{ij}+2E,_{ij}].
 \end{eqnarray}
Here $\psi$ is a perturbation associated to the curvature and E is 
asociated to the expansion. We will work under the Newtonian gauge, which is defined when $B=E=0$. 
An advantage of using this gauge is that here the metric tensor $g_{\mu \nu}$ is diagonal, and so the calculations become much easier.
 We will only work with scalar perturbations, vector and tensor perturbations are eliminated from the beginning,
 so that only scalar perturbations are taken into account. Another advantage in using this gauge is that $\phi$ 
will play the role of the gravitational potential, this will help us to have a simpler physical interpretation, i.e.,
 both potentials $\phi$ and $\psi$ are then related. 
This metric has already been used in other works (see for example \cite{bardeen}, \cite{ma} and \cite{malik}).

The perturbed energy-momentum tensor for the SF can be written as the background value $\mathbf{T}_0\equiv\mathbf{T}_0(t)$ 
plus a perturbation $\delta\mathbf{T}\equiv\delta\mathbf{T}(x^\mu)$ where $x^{\mu}=[t,x^i]$, i. e.
\begin{equation}
\mathbf{T}=\mathbf{T}_0+\delta\mathbf{T}.
\end{equation}

We now derive the perturbed evolution equations for the different quantities mentioned above; the scalar perturbation $\delta\Phi$ 
and the scalar potential $\psi$. For the perturbed energy-momentum tensor, we get
\begin{subequations}
\begin{eqnarray}
\delta T^0_0&=&-\delta\rho_{\Phi}=-(\dot{\Phi}_0\dot{\delta\Phi}-\dot{\Phi}_0^2\psi+V,_{\Phi_{0}}\delta\Phi),\label{rhopert}\\
\delta T^0_i&=&-\frac{1}{a}(\dot{\Phi}_0\delta\Phi,_i),\\
\delta T^i_j&=&\delta p_{\Phi}=(\dot{\Phi}_0\dot{\delta\Phi}-\dot{\Phi}_0^2\psi-V,_{\Phi_{0}}\delta\Phi)\delta^i_j.
\end{eqnarray}\label{eq:Tperturbado}
\end{subequations}
Where in equations (\ref{eq:Tperturbado}) the dot denotes differentiating with respect to cosmological time $t$,
which is related to conformal time by the simple relation
\begin{equation}
\frac{d}{d\eta}=a\frac{d}{dt}.
\end{equation}

In the Newtonian gauge, the metric tensor $g_{\mu\nu}$ becomes diagonal and from this, in the trace of the Einstein's equations, 
the scalar potentials $\psi$ and $\phi$ are identical and, therefore, $\psi$ plays the role of the gravitational potential
\begin{equation}
\psi-\phi=0.
\label{cond}
\end{equation}
Usually this equation contains a term of anisotropic stress, which vanishes in the case of a SF. 
Altogether, the perturbed Einstein's equations $\delta G^i_j= \kappa^{2}\delta T^i_j$ 
to first order for a SF in the Newtonian gauge are
\begin{eqnarray}
-8\pi G\delta\rho_{\Phi}&=&6H(\dot{\phi}+H\phi)-\frac{2}{a^2}\nabla^2\phi, \nonumber\\
8\pi G\dot{\Phi}_0\delta\Phi,_i&=&2(\dot{\phi}+H\phi),_i, \nonumber\\
8\pi G\delta p_{\Phi}&=&2[\ddot{\phi}+4H\dot{\phi}+(2\dot{H}+3H^2)\phi].
\label{eq:sfmet}
\end{eqnarray}
These equations describe the evolution of the scalar perturbations. For the evolution of the perturbations in 
the SF we use the perturbed Klein-Gordon equation
\begin{equation}
\ddot{\delta\Phi}+3H\dot{\delta\Phi}-\frac{1}{a^2}\nabla^2\delta\Phi+V,_{\Phi\Phi}\delta 
\Phi+2V,_{\Phi}\phi-4\dot{\Phi}_0\dot{\phi}=0.
\label{eq:kgl}
\end{equation}

In order to solve the equations (\ref{eq:sfmet}) and (\ref{eq:kgl}), we will turn to Fourier's space. The beauty of this expansion relaying on the fact that each Fourier mode will propagate independently. To first order, the derivation of Fourier's components is straightforward. The perturbation $\delta\Phi$ relates to its Fourier component $\delta\Phi_k$ by
\begin{eqnarray}
\delta\Phi(t,x^i)&=&\int d^3k\delta\Phi(t,k^i)\,\textrm{exp}(ik_ix^i),\nonumber\\&=&\int d^3k\delta\Phi_k\,\textrm{exp}(ik_ix^i),
\end{eqnarray}
where $k$ is the wave number. Here the wave number is defined as $k=2\pi /L$, 
being $L$ the length scale of the perturbation.

The perturbed equations (\ref{eq:sfmet}) altogether with the SF transformation read
\begin{subequations}
\begin{eqnarray}
8\pi G(3H\dot{\Phi}_0\delta\Phi_k)+\frac{2k^2}{a^2}\phi&=&-8\pi G(\dot{\Phi}_0\dot{\delta\Phi_k}\nonumber\\&&-\phi\dot{\Phi}_0^2+V,_{\Phi}\delta\Phi_k),\label{eq:fsa}\nonumber\\ \\
2(H\phi+\dot{\phi})&=&8\pi G\dot{\Phi}_0\delta\Phi_k, \label{eq:fsb}\\
2[\ddot{\phi}+4H\dot{\phi}+(2\dot{H}+3H^2)\phi]&=&
8\pi G(\dot{\Phi}_0\dot{\delta\Phi_k}
\nonumber\\&&-\phi\dot{\Phi}_0^2-V_{\Phi}\delta\Phi_k)\label{eq:fsc},
\end{eqnarray}
\label{eq:fsp}
\end{subequations}
and the Klein-Gordon equation (\ref{eq:kgl}) transforms into
\begin{equation}
\ddot{\delta\Phi}_k+3H\dot{\delta\Phi_k}+\frac{k^2}{a^2}\delta\Phi_k+V,_{\Phi\Phi}\delta\Phi_k-4\dot{\phi}\dot{\Phi}_0+2\phi V,_{\Phi}=0.
\label{eq:kgfou}
\end{equation}
These set of equations describe the evolution of the perturbations. Eq. (\ref{eq:fsa}) makes reference to the evolution
of the energy density, Eq. (\ref{eq:fsb}) to the evolution of the gravitational potential and finally, 
Eq. (\ref{eq:kgfou}) refers to the perturbations over the SF.

\begin{figure}[h]
 \centering
 \scalebox{0.3}{\includegraphics{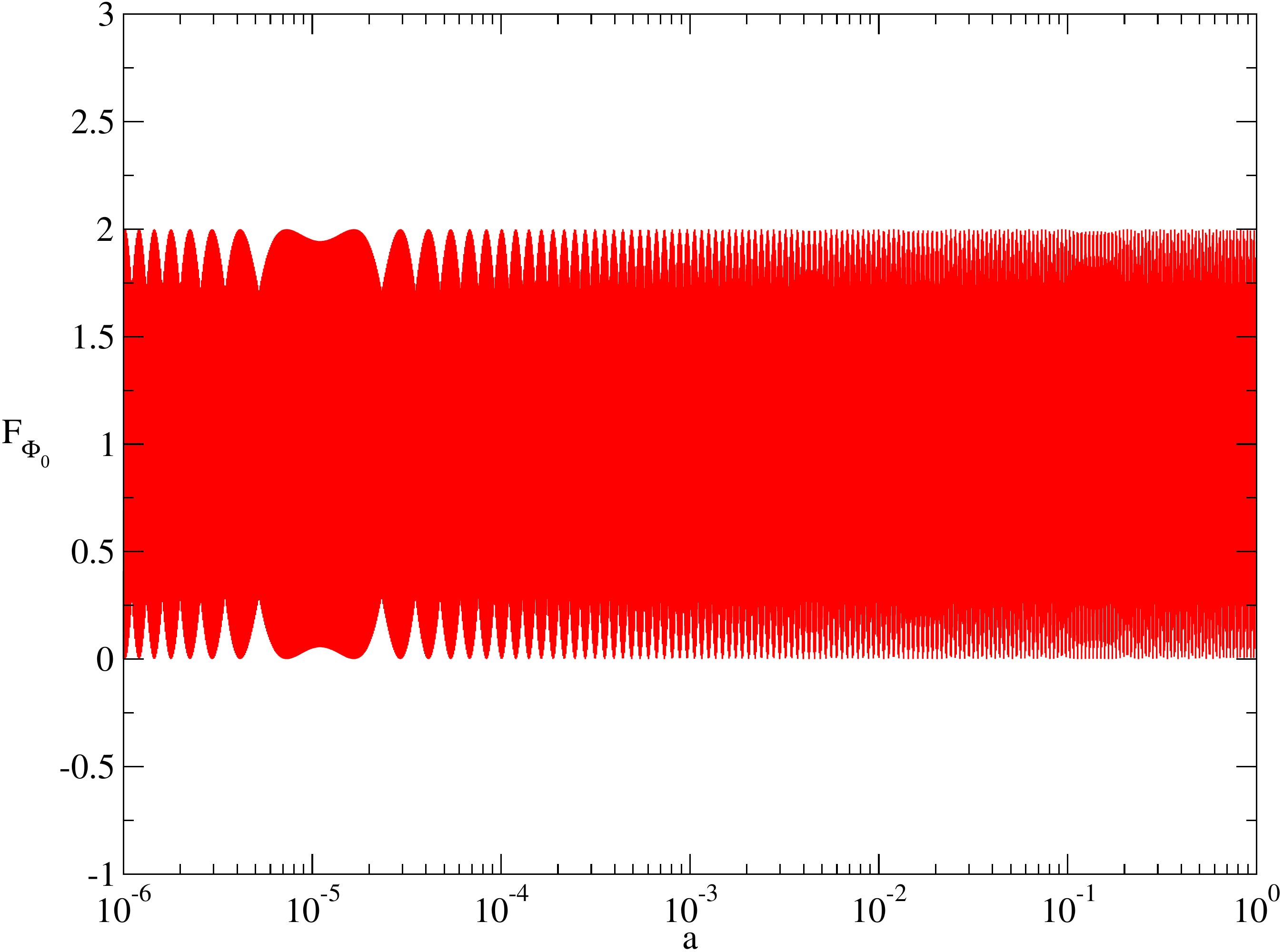}}
 \caption{Evolution of $\left< F_{\Phi} \right>$ term involved on the right-hand side of equation (\ref{eq:deltlini}).}
 \label{fig:FPhi0}
\end{figure}

Now, from the time derivative of (\ref{rhopert}), we have
\begin{eqnarray}
 \dot{\delta\rho_{\Phi}}&=&(\ddot{\Phi}_0+V_{\Phi})\dot{\delta\Phi_k}
 \nonumber\\&&+(\ddot{\delta\Phi_k}+V,_{\Phi\Phi}\delta\Phi_k-\dot{\Phi}_0\dot{\phi})
 \dot{\Phi}_0-2\phi\dot{\Phi}_0\ddot{\Phi}_0.
\end{eqnarray}
Performing a Fourier transformation of the above equation and combining it with the unperturbed 
and perturbed Klein-Gordon equations (\ref{eq:nopertKG}) and (\ref{eq:kgfou}) respectively, 
and with the use of equation (\ref{eq:fsp}) we arrive at
\begin{equation}
\dot{\delta\rho_{\Phi}}=-6H\dot{\Phi}_0\dot{\delta\Phi_k}+6\phi\dot{\Phi}_0^2H-\frac{2k^2}{a^2\kappa^2}(H\phi+\dot{\phi})+3\dot{\phi}
\dot{\Phi}_0^2.
\end{equation}
And on the other hand we have,
\begin{equation}
\delta p_{\Phi}+\delta\rho_{\Phi}=2\dot{\Phi}_0\dot{\delta\Phi_k}-2\dot{\Phi}_0^2\phi,
\end{equation}
Thus,
\begin{equation}
\dot{\delta\rho_{\Phi}}=-3H(\delta p_{\Phi}+\delta\rho_{\Phi})-\frac{2k^2}{a^2\kappa^2}(H\phi+\dot{\phi})+3\dot{\phi}\dot{\Phi}_0^2.
\end{equation}
This last equation can be expressed in terms of the density contrast $\delta_{\Phi}$ 
making use of the equations from the background, we have
\begin{equation}
\dot\delta_{\Phi}+3H(\frac{\delta p_{\Phi}}{\delta\rho_{\Phi}}-\omega_{\Phi})\delta_{\Phi}=3\dot{\phi}(1+\omega_{\Phi})-G_{\phi},
\label{pro1}
\end{equation}
where we have defined the function $G_{\phi}$ as
\begin{equation}
G_{\phi}=\frac{2k^2}{a^2\kappa^2}\frac{\dot\phi+H\phi}{\rho_{\Phi_{0}}}.
\end{equation}
It is convenient to define the function $F_{\phi}$ as well, 
\begin{equation}
F_{\phi}=1+\omega_{\Phi}.
\end{equation}
Taking the time average of equation (\ref{pro1}) we obtain
\begin{equation}
\dot\delta_{\Phi}+3H(\frac{<\delta p_{\Phi}>}{<\delta\rho_{\Phi}>}-<\omega_{\Phi}>)\delta_{\Phi}
=3\dot{\phi}<F_{\Phi}>-<G_{\phi}>.
\label{eq:deltlini}
\end{equation}  
In equation (\ref{eq:deltlini}) for the radiation and matter dominated eras the first term in the parenthesis
goes as $<\delta p_{\Phi}>/< \delta\rho_{\Phi}>\approx 0$, see for example \cite{further}, 
also we can see from Fig.\ref{fig:w0} that $\left<\omega_{\Phi_{0}}\right>\rightarrow 0$. 
It is easy to see the temporal average of the terms $F_{\Phi}$ and $G_{\phi}$ of the equation (\ref{eq:deltlini}). 
From Fig.\ref{fig:FPhi0} we observe that that $\left< F_{\Phi} \right>$ tends to one. 
On the other hand, since we are using post-newtonian approximation $\left<G_{\phi}\right>$ drops to zero, 
meaning that the second term on the right-hand side of Eq. (\ref{eq:deltlini}) dissapears.
So, due to the scalar oscillations around the minimum of the scalar potential, 
the unperturbed SF behaves very similar to the CDM model, but equation (\ref{eq:deltlini}) tell us that their perturbations too, 
all the growing behavior for the $k$ modes are recovered and preserved so far.
Finally, we conclude that Eq. (\ref{eq:deltlini}) resembles the equation for the density contrast as in the standard 
CDM model \cite{chung}. This means that the SFDM perturbations in the linear regime grow exactly as CDM perturbations.

\subsection{The rescaled equations for SFDM perturbations}

We now study the cosmological evolution of our model through a dynamical system. 
In order to solve the system of equations for the perturbations given in Eqs. (\ref{eq:fsp}) and (\ref{eq:kgfou}), 
we define the following dimensionless variables given by
\begin{eqnarray}
U&\equiv&-\frac{\kappa}{\sqrt{6}}\frac{V,_{\Phi_{0}}}{H^2},\,\,\,\,\,\,\,\,\,
l_1\equiv\phi,\nonumber\\
x_2&\equiv&\frac{\dot{\phi}}{H},\,\,\,\,\,\,\,\,\,\,\,\,\,\,\,\,\,\,\,\,\,\,\,\,\,\,\,\,
z_1\equiv\frac{\kappa}{\sqrt{6}}\delta\Phi_k,\nonumber\\
z_2&\equiv&\frac{\kappa}{\sqrt{6}}\frac{\delta\dot{\Phi}_k}{H}.
\label{eq:dvlr}
\end{eqnarray}

Using the variables defined in Eq. (\ref{eq:dvlr}), the equations can be transformed into an autonomus system with 
respect to the e-folding number $N$, then the density contrast can be written in terms of these variables

\begin{equation}
\delta=\frac{2[x(z_2-xl_1)-Uz_1]}{\Omega_{\Phi_{0}}}.
\label{eq:deltala}
\end{equation}

Therefore the transition between the linear and non-linear regime of the SFDM
perturbations can be studied through a numerical code.

\subsection{The Natural Cut of the mass power spectrum}

We rewrite equation (\ref{eq:kgfou}) in the following form

\begin{equation}
\ddot{\delta\Phi}_k+(\frac{k^2}{a^2}+V,_{\Phi\Phi})\delta\Phi_k=-3H\dot{\delta\Phi_k}+4\dot{\phi}\dot{\Phi}_0-2\phi V,_{\Phi}.
\label{eq:kgfou_2}
\end{equation}
Equation (\ref{eq:kgfou_2}) is the one of a harmonic oscillator with a damphing term $3H\dot{\delta\Phi}$ and an extra force. This equation contains oscillating, growing and decaying solutions, depending on the factor $\frac{k^2}{a^2}+V,_{\Phi\Phi}$. If this factor is positive, we expect that the fluctuations oscillate, only if the factor is negative we will have growing up fluctuations. This factor contains the scale $k^2/a^2$, which is always positive. Thus, if we expect to have growing up fluctuation of the scalar field, this happen only when the potential $V,_{\Phi\Phi}$ is negative. Then equation (\ref{eq:kgfou_2}) will have growing up solutions for certain values of the scale $k/a$, this is, only for enough small values of the scale, which means for enough big fluctuations. This is the way we can fix the mass of the scalar field, the potential depends on the value of the scalar field mass. Observing the smallest structure we see in the Universe, we are able to fix the mass of the scalar field. Using this observations one can avoid the problem of substructure of the $\Lambda$CDM. In order to do this, in \cite{further} the mass of the scalar field was fixed in $10^{-22}$eV.

\subsection{The Linear Regime of SFDM perturbations: The Fluid Approach}\label{fluct}

The SF $\Phi$ has very stark oscillations from the beginning, these oscillations are transmitted to the fluctuations 
which apparently seems to grow very fast and are too big. Nevertheless, this behavior is not physical, 
because we only see the oscillations of the fields, but we can not see clearly the evolution of its density, \cite{matos_open}. 
In order to drop out these oscillations, in what follows we perform two transformations. 
The first one changes the perturbed Klein-Gordon equation into a kind of 'Schr\"odinger' equation and the second transforms 
this last equation into a hydrodynamical system, where we can interpret the physical quantities easier and 
the observable quantities become much clear. 

First, we consider the perturbed Klein-Gordon \ref{eq:kgl} where we have set $\dot{\phi}=0$
\begin{equation}
\delta\ddot{\Phi}+3H\delta\dot{\Phi}-\frac{1}{a^2}\hat\nabla^2\delta\Phi+V,_{\Phi\Phi}\delta\Phi+2V,_{\Phi}\phi=0.
\label{K-G}
\end{equation}
Now we express the perturbed SF $\delta\Phi$ in terms of the field $\Psi$,
 \begin{equation}
  \delta\Phi=\Psi \,{e}^{-\,{i}mt/\hbar}+\Psi^*\,{e}^{\,{i}mt/\hbar},
 \end{equation}
term which oscillates with a frequency proportional to $m$ and $\Psi=\Psi(\mbox{\boldmath$x$},t)$ 
which would be proportional to a wave function of an ensamble of particles in the condensate. 
With this equation and the expresion for the potential of the scalar field, (\ref{K-G}) transforms into
   \begin{equation}
    -\,{i}\hbar(\dot{\Psi}+\frac{3}{2}H\Psi)+\frac{\hbar^2}{2m}(\square\Psi+9\lambda|\Psi|^2\Psi)+m\phi\Psi=0,
   \label{schrodinger}
   \end{equation}
where we have defined
 \begin{equation}
  \square=\frac{\,{d}^2}{\,{d}t^2}+3H\frac{\,{d}}{\,{d}t}-\frac{1}{a^2}\hat\nabla^2.
 \end{equation}
Notice that this last equation could represent a kind of 'Gross-Pitaevskii' equation in an expanding Universe. The only modification of equation (\ref{schrodinger}) in comparison to the Schr\"odinger or the Gross-Pitaevskii equation is the scale factor $a^{-1}$ associated to the co-moving spatial gradient and that the Laplacian $\hat\nabla^2=\partial^2_{\mbox{\boldmath{x}}}$ transforms into the D' Alambertian $\Box$.

 To explore the hydrodynamical nature of bosonic dark matter, we will use a modified fluid approach. Then, to make the connection between 
the theory of the field and the condensates waves function, the field is proposed as (Madelung transformation \cite{pitaevskii})
 \begin{equation}
  \Psi=\sqrt{\hat{\rho}}\,\,{e}^{\,{i}S},
 \label{psi}
 \end{equation}
where $\Psi$ will be the condensates wave function with $\hat{\rho}=\hat{\rho}(\mbox{\boldmath$x$},t)$ and $S=S(\mbox{\boldmath$x$},t)$, \cite{ginzburg}. Here we have separated $\Psi$ into a real phase $S$ and a real amplitude $\sqrt{\hat{\rho}}$ and the condition $\mid\Psi\mid^2=\Psi\Psi^*=\hat{\rho}$ is satisfied. From (\ref{psi}) we have
 \begin{eqnarray}
  \dot{\hat{\rho}}&+&3H\hat{\rho}-\frac{\hbar}{m}\hat{\rho}\Box S+\frac{\hbar}{a^2m}\hat\nabla S\hat\nabla\hat{\rho}
   -\frac{\hbar}{m}\dot{\hat{\rho}}\dot{S}=0,\nonumber\\
  \hbar\dot S/m&+&\omega\hat{\rho}+\phi+\frac{\hbar^2}{2m^2}\left(\frac{\Box\sqrt{\hat{\rho}}}{\sqrt{\hat{\rho}}}\right)
  +\frac{\hbar^2}{2a^2}[\hat\nabla(S/m)]^2\nonumber\\ 
  &-&\frac{\hbar^2}{2}(\dot S/m)^2=0.\label{hidro1}
 \label{hidro}
 \end{eqnarray}
Now, taking the gradient of (\ref{hidro1}) then dividing by $a$ and using the definition
 \begin{equation}
  \mbox{\boldmath$v$}\equiv\frac{\hbar}{ma}\hat\nabla S
 \label{vel}
 \end{equation}
we have,
 \begin{eqnarray}
  \dot{\hat{\rho}}&+&3H\hat{\rho}-\frac{\hbar}{m}\hat{\rho}\Box S+\frac{1}{a}\mbox{\boldmath$v$}\nabla\hat{\rho}
   -\frac{\hbar}{m}\dot{\hat{\rho}}\dot{S}=0,\nonumber\\
  \dot{\mbox{\boldmath$v$}}&+&H\mbox{\boldmath$v$}+\frac{1}{2a\hat{\rho}}\nabla p+\frac{1}{a}\nabla\phi
  +\frac{\hbar^2}{2m^2a}\nabla\left(\frac{\Box\sqrt{\hat{\rho}}}{\sqrt{\hat{\rho}}}\right)\nonumber\\
  &+&\frac{1}{a}(\mbox{\boldmath$v\cdot$}\nabla)\mbox{\boldmath$v$}-\hbar(\dot{\mbox{\boldmath$v$}}+H\mbox{\boldmath$v$})
  (\dot S/m)=0\label{eq:1}.
 \end{eqnarray}
where in (\ref{hidro1}) $\omega=9\hbar^2\lambda/2m^2$ and in (\ref{eq:1}) we have defined $p=\omega\hat{\rho}^2$.

  It is worth noting that, to this moment this last set of equations do not involve any approximations with respect to equation 
(\ref{schrodinger}) and can be used in linear and non-linear regimes.

 Now, neglecting squared terms, second order time derivatives and products of time derivatives in this last set of equations we get,
 \begin{eqnarray}
  \frac{\partial\hat{\rho}}{\partial t}&+&\nabla\mbox{\boldmath$\cdot$} (\hat{\rho}\mbox{\boldmath$v$})+3H\hat{\rho}=0\label{navier},\\
  \frac{\partial\mbox{\boldmath$v$}}{\partial t}&+&H\mbox{\boldmath$v$}+(\mbox{\boldmath$v\cdot$}\nabla)\mbox{\boldmath$v$}- 
  \frac{\hbar^2}{2m^2}\nabla(\frac{1}{2\hat{\rho}}\nabla^2\hat{\rho})+\omega\nabla\hat{\rho}\nonumber\\
  &+&\nabla\phi=0\label{navier1},\\
  \nabla^2\phi&=&4\pi G\hat{\rho},\label{poisson}
 \end{eqnarray}
where the equation for the gravitational field is given by the Poisson's equation (\ref{poisson}). 
In these equations we have introduced $\mbox{\boldmath$r$}=a(t)\mbox{\boldmath$x$}$, such that 
$1/a\hat\nabla=\nabla=\partial_{\mbox{\boldmath{r}}}$.

Equation (\ref{vel}) shows the proportionality between the gradient of the phase and the velocity of the fluid. 
Note that {\boldmath$v$} can represent the velocity field for the fluid and $\hat{\rho}$ will be the particles density number 
within the fluid. Also there exists an extra term of third order for the partial derivatives in the waves 
amplitude which goes as the gradient of $\frac{\hbar^2}{2m^2}\frac{\Box\sqrt{\hat{\rho}}}{\sqrt{\hat{\rho}}}$, 
this term would result in a sort of 'quantum pressure' that would act against gravity. 
We remain that $\phi$ represents the gravitational field. These two sets of equations (\ref{navier}) and (\ref{navier1}) 
would be analogous to the Euler's equations of classical 'fluids', with the main difference that there exists a 'quantum term', 
which we will call $Q$ and will be given by $Q=\frac{\hbar^2}{2m^2}\frac{\square\sqrt{\hat{\rho}}}{\sqrt{\hat{\rho}}}$ 
which can describe a force or a sort of negative quantum pressure.

For equation (\ref{navier}) we have that $\hat{\rho}$ will represent the mass density or the particles density number of the fluid, where 
all the particles would have the same mass. Finally, these equations describe the dynamics of a great number of non-interacting identical
particles that manifest themselves in the form of a fluid, also equation (\ref{schrodinger}) can describe a great 
number of non-interacting but self-interacting identical particles in the way of a Bose gas, 
when the probability density is interpreted as the density number.

Now, these hydrodynamical equations are a set of complicated non-linear differential equations. To solve them we will restrict ourselves to 
a vecinity of total equilibrium.

 For this let $\hat{\rho}_0$ be the mass density of the fluid in equilibrium, the average velocity 
$\mbox{\boldmath$v$}_0$ will be taken as zero in equilibrium, so we will only have $\mbox{\boldmath$v$}(\mbox{\boldmath$x$},t)$ out of equilibrium. Then, the matter in the Universe will be considered as a hydrodynamical fluid inside an Universe in expansion. This system will then evolve in this Universe and later on they will collapse because of their gravitational attraction.

   Then from (\ref{navier}) for the mass density of the fluid in equilibrium we have,
   \begin{equation}
    \frac{\partial\hat{\rho}_0}{\partial t}+3H\hat{\rho}_0=0,
   \end{equation}
with solutions of the form
   \begin{equation}
    \hat{\rho}_0=\frac{\rho_{0i}}{a^3},
   \label{back}
   \end{equation}
where as we know, in general if we have an equation of state of the form $\hat p=\omega\hat\rho$ and consider CDM or 
dust as dark matter such that $\hat p=0$ it holds that $\hat\rho\propto a^{-3}.$ Then, when the scale factor was small, 
the densities were necessarily bigger. Now, the particles density number are inversely proportional to the volume, 
and must be proportional to $a^{-3}$, therefore the matter energy density will also be proportional to $a^{-3}$, 
result that is consistent with our expression (\ref{back}). In addition, the cosmological behavior of $\hat{\rho}_0$
is in agreement with the numerical results show in Fig. \ref{fig:back}.

Now for the system out of equilibrium we have
 \begin{eqnarray}
  \frac{\partial\delta\hat{\rho}}{\partial t}&+&3H\delta\hat{\rho} 
  +\hat{\rho}_0\nabla\mbox{\boldmath$\cdot$}\delta\mbox{\boldmath$v$}=0,\nonumber\\
  \frac{\partial\delta\mbox{\boldmath$v$}}{\partial t}&+&H\delta\mbox{\boldmath$v$} 
  -\frac{\hbar^2}{2m^2}\nabla(\frac{1}{2}\nabla^2\frac{\delta\hat{\rho}}{\hat{\rho}_0})+\omega\nabla\delta\hat{\rho}
  +\nabla\delta\phi=0,\nonumber\\
  \nabla^2\delta\phi&=&4\pi G\delta\hat{\rho},
  \label{outeq}
  \end{eqnarray}
equations that are valid in a Universe in expansion. In order to solve system (\ref{outeq}) we look for solutions in the form of plane waves, for this the convenient ansatz goes as 
 \begin{eqnarray}
  \delta{\hat\rho}&=&\hat\rho_1(t)\exp(\,{i}\mbox{\boldmath$k\cdot x$}/a),\nonumber\\ 
  \delta\mbox{\boldmath$v$}&=&\mbox{\boldmath$v$}_1(t)\exp(\,{i}\mbox{\boldmath$k\cdot x$}/a),\nonumber\\
  \delta\phi&=& \phi_1(t)\exp(\,{i}\mbox{\boldmath$k\cdot x$}/a).\nonumber
 \end{eqnarray}
where {\boldmath$x$} is the position vector and {\boldmath$k$} is a real wavevector which corresponds to a wavelength $L$. If we substitute these ansatz in the set of equations (\ref{outeq}), we then have
   \begin{eqnarray}
   \frac{\,{d}\hat\rho_1}{\,{d} t}&+&3H\hat\rho_1+\,{i}\frac{{\hat\rho}_0}{a}\mbox{\boldmath$k\cdot v$}_1=0,\label{eq1}\\
   \frac{\,{d}\mbox{\boldmath$v$}_1}{\,{d} t}&+&H\mbox{\boldmath$v$}_1+\,{i}\frac{\hat\rho_1}{a}\left(\frac{v_{q}^2}{\hat{\rho}_0}-
     4\pi G\frac{a^2}{k^2}+\omega\right)\mbox{\boldmath$k$}=0,\\
    \phi_1&+&4\pi G\frac{a^2}{k^2}\hat\rho_1=0,
   \label{modos}
   \end{eqnarray}
where we have defined the velocity
\begin{equation}
v^2_{q}=\frac{\hbar^2k^2}{4a^2m^2}.
\label{eq:vq}
\end{equation}
   To solve the system is convenient to rotate the coordinate system so that the propagation of the waves will be along the direction of 
one of the axes. For this we know that the velocity vector can be divided into longitudinal (parallel to {\boldmath$k$}) and transverse (perpendicular to {\boldmath$k$}) parts, such that $\mbox{\boldmath$v$}_1=\nu\mbox{\boldmath$k$}+\mbox{\boldmath$v$}_2$, where $\mbox{\boldmath$v$}_2$ is the vector perpendicular to the wave propagation vector $\mbox{\boldmath$k\cdot v$}_2=0$. In terms of $\mbox{\boldmath$v$}_2$ for equations (\ref{eq1})-(\ref{modos}) we have
   \begin{eqnarray}
   \frac{\,{d}\hat\rho_1}{\,{d} t}&+&3H\hat\rho_1+\,{i}\frac{{\hat\rho}_0}{a}k^2\nu=0,\label{eq2}\\
    \frac{\,{d}\nu}{\,{d} t}&+&H\nu+\frac{\,{i}}{a}(\frac{v_{q}^2}{\hat{\rho}_0}-4\pi G\frac{a^2}{k^2}+\omega)\hat\rho_1=0,
    \label{lambda}
   \end{eqnarray}
in addition to an equation for $\mbox{\boldmath$v$}_2$, $\,{d}\mbox{\boldmath$v$}_2/\,{d} t+H\mbox{\boldmath$v$}_2=0$, with solutions $\mbox{\boldmath$v$}_2=C/a$ with $C$ a constant of integration, i.e., perpendicular modes to the wave vector are eliminated with the expansion of the Universe. Now, if we use the result (\ref{back}), then equation (\ref{eq2}) can be written as
   \begin{equation}
    \frac{\,{d}}{\,{d}t}\left(\frac{\hat\rho_1}{\hat{\rho}_0}\right)=-\frac{\,{i}k^2\nu}{a}.   \label{lambda1}
   \end{equation}
 System (\ref{eq2})-(\ref{lambda}) can be treated as in the case of a Universe with no expansion, so combining the two equations and with 
the aid of (\ref{lambda1}), we get
   \begin{equation}
    \frac{\,{d}^2\delta}{\,{d}t^2}+2H\frac{\,{d}\delta}{\,{d}t}+\left[(v_{q}^2+\omega\hat{\rho}_0)\frac{k^2}{a^2}
    -4\pi G{\rho}_0\right]\delta=0,
   \label{delta}
   \end{equation}
where $\delta=\hat\rho_1/\hat{\rho}_0=\rho_1/{\rho}_0$ is defined as the density contrast. This will be a fundamental equation in the understanding of the evolution of the primordial fluctuations.

\section{Results}\label{results}

First we will give a brief summary of the results for the $\Lambda$CDM model, this will enable us to make a direct comparison with our results.

 For CDM the equation for the evolution of the density contrast is given by,
 \begin{equation}
  \frac{\,{d}^2\delta}{\,{d}t^2}+2\,H\frac{\,{d}\delta}{\,{d}t}+\left(c_{s}^2\frac{k^2}{a^2}-4\pi G\hat{\rho}_0\right)\delta=0,
 \label{deltaCDM}
 \end{equation}
where $c_{s}$ is defined as the sound velocity (which in our case it is not). Now lets analyze equation (\ref{deltaCDM}) at the beginning of the matter dominated era a time just after the epoch of equality, and just before recombination when the radiation has cooled down and the photons do not interact with the electrons anymore, for a relativistic treatment see \cite{gorini}. In this era, $a\geqslant a_{eq}$, practically all the interesting fluctuation modes are well within the horizon, and the evolution of the perturbations can be well described within the newtonian analysis. At this time, matter behaves like dust with zero pressure. So we have $a\sim t^{2/3}$, $c_{s}^2k^2/a^2\approx 0$ and $\hat{\rho}_0\sim t^{-2}$ therefore $H=(2/3)1/t$. For equation (\ref{deltaCDM}) we have
 \begin{equation}
  \frac{\,{d}^2\delta}{\,{d}t^2}+\frac{4}{3}\frac{1}{t}\frac{\,{d}\delta}{\,{d}t}-\frac{2}{3}\frac{1}{t^2}\delta=0.
 \end{equation}

   The solutions to this equation are of the form
   \begin{equation}
    \delta(t)\to t^{2/3}C_1+\frac{C_2}{t},
   \label{deltat}
   \end{equation}
where $C_1$ and $C_2$ are integration constants, from this solution we can see that we have modes that will disappear as time goes by, and modes that grow proportionally to the expansion of the Universe. This is an important result, because then the density contrast will grow proportionally to the expansion of the Universe when this is dominated by matter. Then, these fluctuations can maybe grow and give life to the galaxies, clusters of galaxies and all the large-scale structure we see now a days.
\begin{figure}
%\vspace{174pt}
 \scalebox{0.6}{\includegraphics{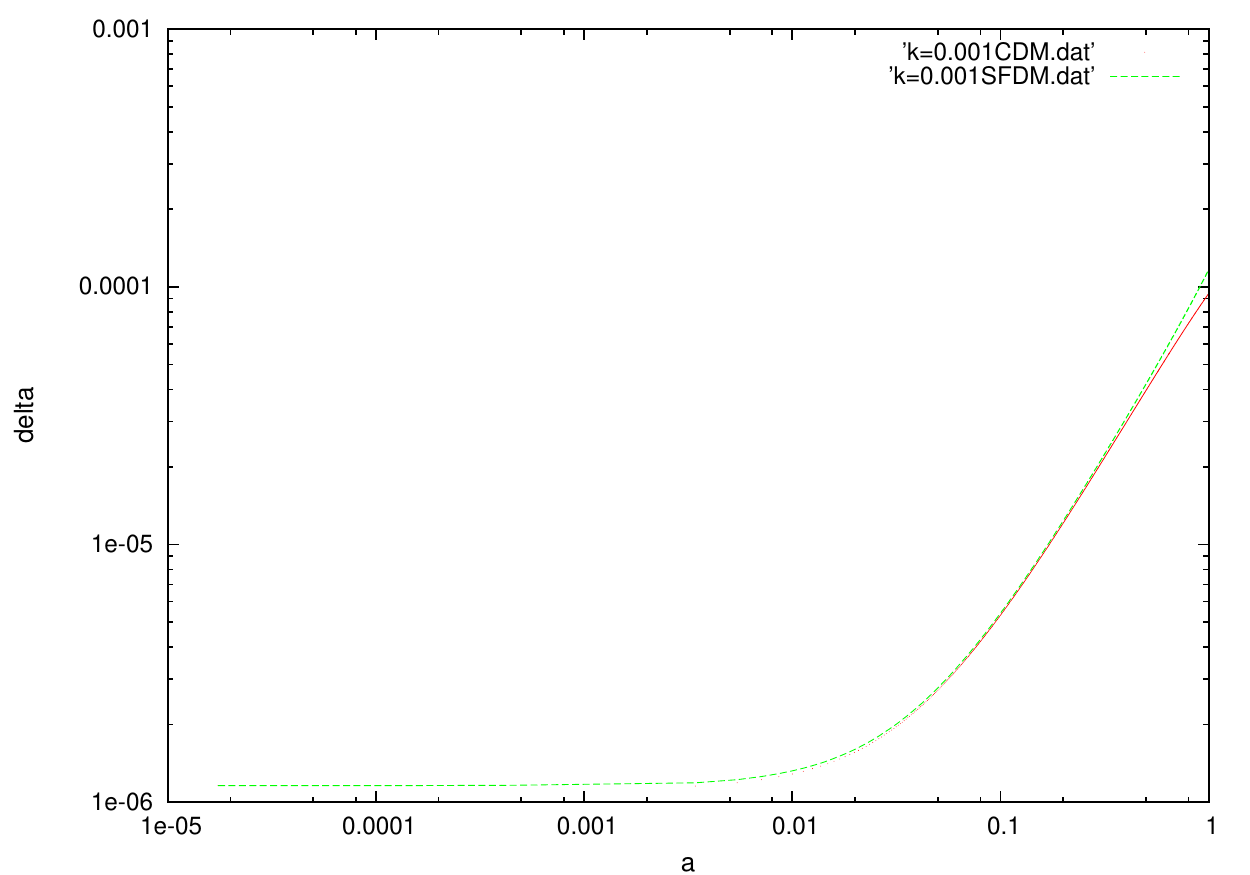}}
\caption{Evolution of the perturbations for the CDM model (dots) and SFDM model (lines) for $k=1*10^{-3}hMpc^{-1}$. Notice how after the 
epoch of equality ($a_{eq}\sim 10^{-4}$) the evolution of both perturbations in nearly identical, $a=1$ today. In this case we have taken $\lambda=0$.}
 \label{fig2}
 \end{figure}

 Now lets see what happens to the SFDM at this epoch ($a\geqslant a_{eq}$). The evolution of the perturbations in this case will 
be given by equation (\ref{delta}).

 In general we have that in equation (\ref{delta}) the term $v_q$ is very small throughout the evolution of the pertubations ($v_q\leq 
10^{-3}ms^{-1}$ for small $k$), so it really does not have a significant contribution on its evolution.
 
 When the condition $\lambda=0$ is taken we can have a BEC that might be or might not be stable, if there exists stability the results of 
SFDM are consistent with those obtained from CDM (in this case both equations (\ref{delta}) and (\ref{deltaCDM}) are almost equal), the condition of stability for the BEC in the SFDM case will come from the study of $\lambda$ together with $Q$. 
\begin{equation}
    \frac{\,{d}^2\delta}{\,{d}t^2}+2H\frac{\,{d}\delta}{\,{d}t}+\left(v_{q}^2\frac{k^2}{a^2}
    -4\pi G{\rho}_0\right)\delta=0,
   \label{deltal}
   \end{equation}

 As we can see in Fig. \ref{fig2} the perturbations used for the $\Lambda$CDM model grow in a similar way for the SFDM model, 
when $\lambda=0$, in this case both perturbations can give birth to structures quite similar in size, and this will happen with all the fluctuations as long as $k$ is kept small.

\begin{figure}
%\vspace{174pt}
 \scalebox{0.6}{\includegraphics{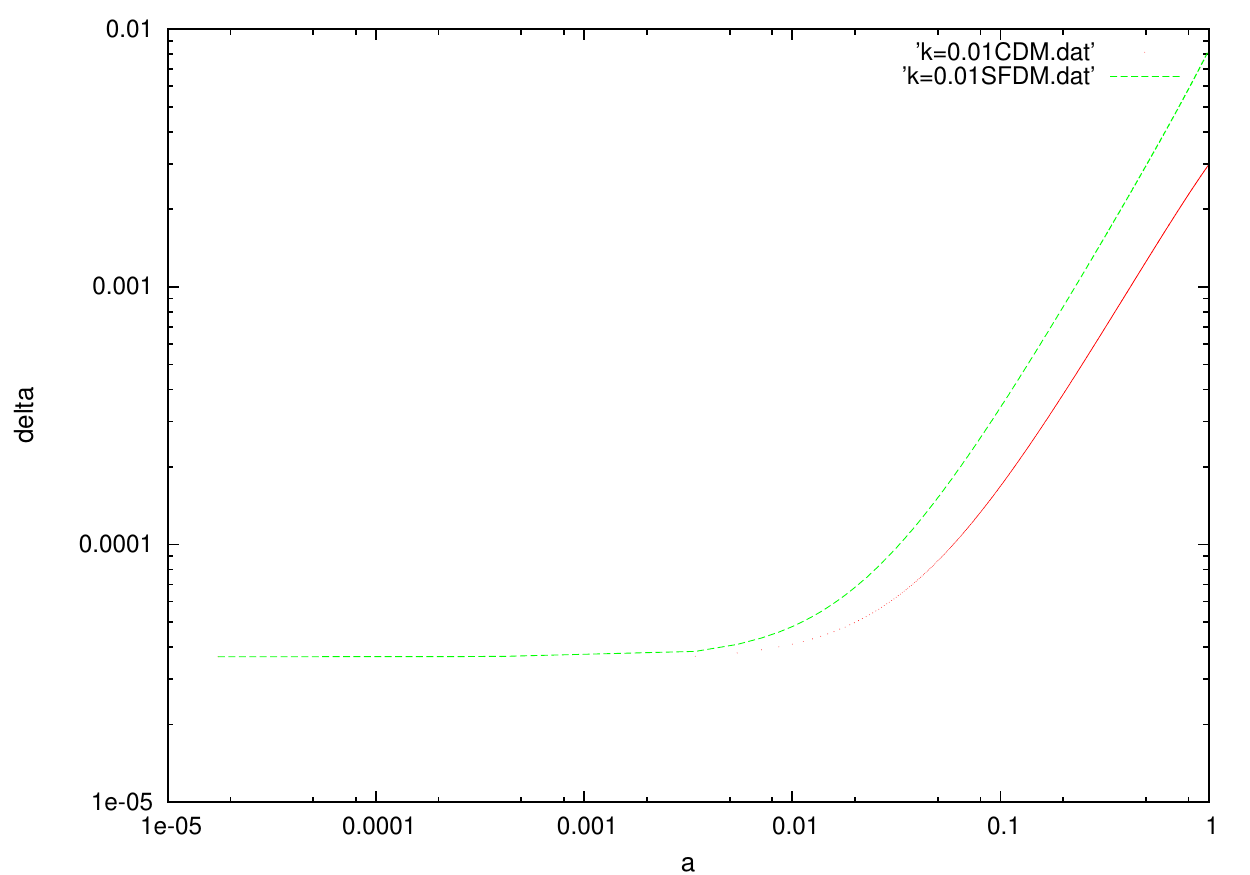}}
  \caption{Evolution of the perturbations for the CDM model (dots) and SFDM model (lines) for $k=1*10^{-2}hMpc^{-1}$ and $\lambda\neq 0$ 
and negative. Notice how after the epoch of equality ($a_{eq}\sim 10^{-4}$) the evolution of both perturbations is now different from the one in Fig. \ref{fig2}, $a=1$ today. In this case we can clearly see that the SFDM fluctuations grow quicker than those for the CDM model.}
 \label{fig3}
 \end{figure}

 When $\lambda\neq 0$ the results are quite different, so when discussing the evolution of the density perturbations, there are two 
different cases: i) In the case of $\lambda >0$ the amplitude of the density contrast tends to decrease as $\lambda$ grows bigger and bigger away from zero until the amplitude of the density takes negative values (around $\lambda\sim 10^{8}$), telling us that this kind of fluctuations can not grow in time, and hence do not form a BEC. ii) On the other hand if $\lambda <0$ the fluctuations for the density contrast alway grow despite their size, this results means either than the fluctuations grow and form a stable BEC or than the density grows because it is collapsing into a single 
point and our BEC might be unstable, the study of the stability of this fluctuations needs then to be studied with non-linear perturbation theory. These results are shown in Fig. \ref{fig3}, in both figures \ref{fig2} and \ref{fig3} the initial condition for $\delta$ goes as $\delta\sim 1*10^{-5}$ in accordance with the data obtained from WMAP.

If these fluctuations result stable and because they are big in size, this means that they can only give birth to large structures. These fluctuations can then help for the formation of large clusters or other large-scale structure in the Universe at its early stages (around $a\geqslant a_{eq}$). Then, as these kind of SFDM can only interact with radiation in a gravitational form it is not limited by its interaction with radiation, and the dark matter halos can then create potential wells that will collapse early in time giving enough time for the structures to form. Then if DM is some kind of SFDM, the luminous matter will follow the DM potentials giving birth to large-scale structure.

Finally, a further analysis is required to obtain better results of
the growth of scalar perturbations with self-interacting SFDM as well as to study
the transition between linear and no linear regime of scalar fluctuations.
Nevertheless, we conclude that the SFDM model at cosmological scale
is a plausible alternative to DM nature because it preserves
all the success of the $\Lambda$CDM model. 

\section{Scalar Field Dark Matter in galactic dynamics}\label{galacticSFDM}

%\subsection{Flat Central Density Profile}
In this section we study the SFDM/BEC model at galactic scales, we will focus on the
cusp/core discrepancy between the numerical predictions of the standard model and
the astronomical observations. 

As already said before, recent observations in far and nearby galaxies have shown that the $\Lambda$CDM model 
faces serious conflicts when trying to explain the galaxy formation at small scales \citep{victor,friedmann}.
For instance, in the $\Lambda$CDM simulations the halos present rising densities towards the central region 
behaving as $\rho \sim r^{-1}$ well within 1 kpc \citep{navarro2010}. On the other hand, several 
observations suggest that the dynamics (rotation curves) of dwarf and LSB galaxies are more consistent with a constant 
central density \citep{kuzio,deblok1, sanchez1}, this is most commonly known as the cusp/core problem \citep{deblok1}.

Studying a wide range of galaxies of different morphologies and with magnitudes in the interval $-22 \leq M_{B} \leq -8$ 
\citet{donato} fit their rotation curves (RC) using a Burkert profile for the DM \citep{burkert} 
and found that 
\begin{equation}
\log (\mu_{0}/M_{\odot}pc^{-2})= 2.15 \pm 0.2 \label{eq:logmu}
\end{equation}
remains approximately constant, where 
\begin{equation}
\mu_{0} = \rho_{0}r_{0} \label{eq:mu}
\end{equation}
with $\rho_{0}$ the central DM
density and $r_{0}$ the core radius. Similar result where found in \citet{kormendy,spano}. 
Exploring further the constant value of $\mu_{0}$ for the DM, \citet{gentile} found that within $r_{0}$ 
the DM central surface density in terms of the mass inside it ($M_{< r_{0}}$) is $<\Sigma>_{0,DM} = M_{< r_{0}} / \pi r^{2}_{0}
\approx 72^{+42}_{-27} $M$_{\odot}$ pc$^{-2}$,the gravitational acceleration due to DM felt by a test particle at the radius $r_{0}$
was found to be 
\begin{equation}
g_{DM}(r_{0})=G \pi <\Sigma>_{0,DM} = 3.2^{+1.8}_{-1.2}\times 10^{-9} \textrm{cm s}^{-2}, \label{eq:gDM}
\end{equation}
additionally they reported the acceleration due to the luminous matter at
$r_{0}$ to be $g_{bar}(r_{0})= 5.7^{+3.8}_{-2.8}\times 10^{-10} \textrm{cm s}^{-2}$.

In the $\Lambda$CDM model the galaxies have evolve through numerous mergers and grew in different environments, the 
star formation and basic properties of the galaxies are not expected to be a common factor among them. Therefore 
giving both the constancy of $\mu_{0}$ and the core in the central regions of galaxies seems very unlikely in this model.

Any model trying to become a serious alternative to $\Lambda$CDM has to succeed in reproducing 
observations in which the standard model fails but also has to keep the solid description at large scale.
It is always necesary to test the SFDM/BEC model with the two observations mentioned above, 
the cusp/core problem and the constant DM central surface density. In order to do this, 
in \cite{victor} it was used the Thomas-Fermi approximation and a static BEC DM halo to fit rotation curves of a set of galaxies. 
However, so far there was no comparison between the density profile 
and the data, in that work this was done by fitting rotation curves of 13 high resolution 
low surface brightness (LSB) galaxies and additionally compare the fits to two
characteristic density profiles 1) the cuspy Navarro-Frenk-White (NFW) profile that results from N-body simulations 
using $\Lambda$CDM and 2) the Pseudo Isothermal (PI) core profile. The comparison allow to show 
that the model is in general agreement with the data and with a core in the central region. 
For the second goal the meaning of a core is somewhat ambiguous. 

Large scale $N$-body simulations using collissionless CDM the internal 
region of DM halos show a density distribution described by a power law $ \rho \sim r^{\alpha}$ with $\alpha \approx -1$,
such behaviour is what is now called a cusp. On the other hand, observations mainly in dwarf and LSB galaxies 
seem to prefer a central density going as $ \rho \sim r^{0}$. 
This discrepancy between observation and the CDM model receives the name of cusp/core problem. 
Among the empirical profiles most frequently used to describe the constant density behavior 
in this galaxies are the PI \citep{begeman}, the isothermal \citep{athana} and the Burkert
profile \citep{burkert}. Even though their behavior is similar in the central region and is specified by the 
central density fitting parameter, their second parameter called the core radius  
does not represent the same idea. For instance, in the PI profile (eq.(\ref{eq:PI})) we see that the core radius will be the 
distance in which the density is half the central density. For the Burkert profile the core radius $R^{burk}_{c}$ 
will be when $\rho^{burk}(R^{burk}_{c}) = \rho^{burk}_{0}/4$ and for an isothermal profile (I) \citep{spano} 
$\rho^{I}(R^{I}_{c}) = \rho^{I}_{0}/2^{3/2}$. Hence, we see an ambiguity in the meaning of the core radius, 
they get the same name but the interpretation depends on the profile. If we want 
to compare the central density of LSB galaxies with that of NFW, it usually suffice to have a 
qualitative comparison, so far this is what we have been doing by fitting 
empirical profiles. However, high resolution rotation curves demand a more quantitative 
comparison. Indeed, if we want to test models by fitting RCs we have to know the specific 
meaning and size of the core, then we will be able to tell if a model is consistent with 
a cusp or not by making a direct comparison with the data. 

In \cite{victor} it was proposed a new 
definition for the core and core radius that allow us to decide when a density profile is cusp or core.
Using this definition in the SFDM/BEC model discussed above it is possible to find that the SFDM/BEC model can reproduced the constant 
value of $\mu_{0}$ and as a crosscheck with the PI profile, in \cite{victor} was found that the results are in very good agreement with observations.
This argues in favor of the model and the core definition.

\subsection[]{DM density profiles}
Following \cite{victor} we provide the dark matter density profiles that will be used for the analysis. 
%In the last part we briefly describe the usual meaning of the core and 
%establish a new definition for it.

The case in which the dark matter is in the form of a static BEC 
and the number of DM particles in the ground state is very large was considered in \citet{boharko}.
Following this paper and assuming the Thomas-Fermi approximation \citep{dalfovo, pitaevskii} which 
neglects the anisotropic pressure terms that are relevant only in the boundary of the condensate,
the system of equations describing the static BEC in a gravitational potential $V$ is given by
\begin{equation}
\nabla p \biggl( \dfrac{ \rho}{m} \biggr) = - \rho \nabla V, \label{eq:BEC1}
\end{equation}

\begin{equation}
 \nabla^{2} V = 4 \pi G \rho, \label{eq:BEC2}
\end{equation}
with the following EoS
\begin{equation}
p(\rho) = U_{0} \rho^{2},
\end{equation}
where $ U_{0} = \dfrac{2 \pi \hbar^{2} a}{m^{3}}$, $\rho$ is the mass density of the static BEC 
configuration and $p$ is the pressure, as we are considering zero temperature $p$ is not a thermal pressure
but instead it is produced by the strong repulsive interaction between the ground state bosons. 
Assuming spherical symmetry and denoting $R$ as the radius at which the pressure and density are zero, 
the density profile takes the form \citep{boharko} 
\begin{equation}
 \rho_{B} (r) = \rho^{B}_{0} \dfrac{\sin( k r)}{k r} \label{eq:BECrho}
\end{equation}
where $k = \sqrt{G m^{3} / \hbar^{2} a}=\pi/R$ and $ \rho^{B}_{0}=\rho_{B}(0)$ 
is the BEC central density, $m$ is the mass of the DM particle and $a$ is the 
scattering lenght. The mass at the radius $r$ is given by 
\begin{equation}
m(r)= \frac{4 \pi \rho^{B}_{0}}{k^{2}} r \biggl( \dfrac{\sin (kr)}{kr } - \cos(kr) \biggr ), \label{eq:BECm}
\end{equation}
from here the tangential velocity $V$ of a test particle at a distance $r$, is  
\begin{equation}
V^{2}(r) = \dfrac{4 \pi G \rho^{B}_{0} }{k^{2}} \biggl( \dfrac{\sin (kr)}{kr } - \cos(kr) \biggr ). \label{eq:BECV}
\end{equation}
The logaritmic slope of a density profile is defined as 
\begin{equation}
\alpha = \dfrac{d (\log \ \rho)}{d (\log \ r)} 
\label{eq:alpha}
\end{equation}
using (\ref{eq:BECrho}) in (\ref{eq:alpha}) we obtain 
\begin{equation}
\alpha(r)= - \biggl [1- \dfrac{\pi r}{R} \cot \biggl(\dfrac{\pi r}{R} \biggr)  \biggr].
\end{equation}

We study now the Pseudo Isothermal (PI) profile. The empirical core profiles that exist in the literature fit two parameters, a scale
radius and a scale density. A characteristic profile of this type is
\begin{equation}
\rho_{PI} = \dfrac{\rho^{PI}_{0}}{1 \ + \ (r/R_{c})^{2}},\label{eq:PI}
\end{equation} 
this is the PI profile \citep{begeman}. Here $R_{c}$ is the scale radius and $\rho^{PI}_{0}$ is the central 
density. The rotation curve is 
\begin{equation}
V(r)_{PI} = \sqrt{4 \pi G \rho^{PI}_{0} R^{2}_{c} \biggl ( 1-\frac{R_{c}}{r} \arctan \biggl ( \dfrac{r}{R_{c}} \biggr) \biggr ) }.
\end{equation}

The NFW profile emerges from numerical simulations that use only CDM and are based on the $\Lambda$CDM model \citep{dubinski,navarro96,navarro97}. 
In addition to this, we have chosen this profile because it is representative of what is called 
the cuspy behavior ($\alpha \approx -1$) in the center of galaxies due to DM. 
The NFW density profile\citep{navarro97} and the rotation curve are given respectively by
\begin{equation}
\rho_{NFW}(r) = \dfrac{\rho_{i}}{(r/Rs)(1 \ + \ r/R_{s})^{2}} 
 \end{equation}
\begin{equation}
V_{NFW}(r)= \sqrt{4 \pi G \rho_{i} R^{3}_{s}} \sqrt{ \frac{1}{r} \biggl [ ln \biggl (1 \ + \ \dfrac{r}{R_{s}} \biggr ) - 
\dfrac{ r/R_{s}}{1\ + \ r/R_{s} } \biggr ]},
\end{equation}
$\rho_{i}$ is related with the density of the universe at the moment the halo collapsed and $R_{s}^{2}$ 
is a characteristic radius.

To solve the core radius ambiguity and to unify the concept for future comparison, in \cite{victor} it was found that a good 
definition for the $core$ is a region where the density profile presents logarithmic slopes $\alpha \geq -1$ and 
the core radius will be the radius at which the core begins, that is to say, for radius smaller 
than the core radius we will have $\alpha \geq -1$, this means that its value $r'$ is determined by the equation
\begin{equation}
\alpha(R) = -1.
\end{equation}
The advantages of this definition are that the interpretation is independent of the profile 
chosen (also notice that it applies to the total density profile and is 
not restricted to that of DM) and in virtue of the same definition 
we can directly tell if a DM model profile is cored or cuspy.  
With this new definition the specific distance at which the core radius occurs still 
depends on the profile chosen but now the physical interpretation is only one. 
In the following when we refer to both the core and core radius we adopt the previous interpretation. 

Applying the definition to (\ref{eq:BECrho}) we get the core radius for the SFDM/BEC profile $R_{B}$ and 
for comparison we use (\ref{eq:PI}) because in turns out that the parameter $R_{c}$ 
corresponds to the core radius as defined above. Finally, fitting the NFW profile 
provides a direct comparison between a cusp and core and hence to the cusp/core problem.

\subsection[]{SFDM vs. PI and NFW}

We see from (\ref{eq:BECrho}) that the SFDM/BEC model satisfies $\rho \sim r^{0}$ near the origin, but a priori 
this does not imply consistency with observed RCs. 
Therefore we reproduce the fit of the profiles to thirteen high resolution observed RCs of a sample of LSB galaxies given in \cite{victor}.
The RCs were taken from a subsample of \citet{deblok2}, we chose galaxies that 
have at least 3 values within $\sim 1$ kpc, not presenting bulbs and the quality in the RC in H$\alpha$ 
is good as defined in \citet{mcgaugh}. The RCs in this work omit 
galaxies presenting high asymmetries and included in the error bars are experimental
errors in the velocity measurement, inclination and small asymmetries.
Because the DM is the dominant mass component for these galaxies 
we adopt the minimum disk hypothesis (neglects baryon contribution to the observed RC). 

As the difference between a core and a cusp is not overlapped only for data values inside $1$kpc
and given that in the interval $\sim 1$ to $10$kpc the slopes of core and cusp profiles are very similar, 
which can lead to the wrong conclusion that cuspy halos are consistent with observations, we 
determined the logarithmic slope and the uncertainty following \citet{deblok2} with 
the difference that we fit only the data within $1$ kpc and that there is no need of an uncertain ``break radius''. 

In Fig. \ref{fig:rotcur}, we show the fits to the RC data and the density profiles, also shown are 
the core radius in the BEC (magenta) and in PI (blue) profiles. The gray arrow is the fit
that determines $\alpha$, the length denotes the fitted region and is bounded by $R_{1}$ that
denotes the nearest radius to $1$kpc where a data point is given. We show $\alpha$ by
fitting values inside $R_{1}$ and we also show the core radius for the SFDM/BEC profile $R_{B}$ in order to compare it with $R_{c}$ 
(see \cite{victor} for more details of the fitting parameters). 

\begin{figure*}
\begin{minipage}{170mm}
\begin{tabular}{@{}lll@{}}
 \resizebox{50mm}{!}{\includegraphics{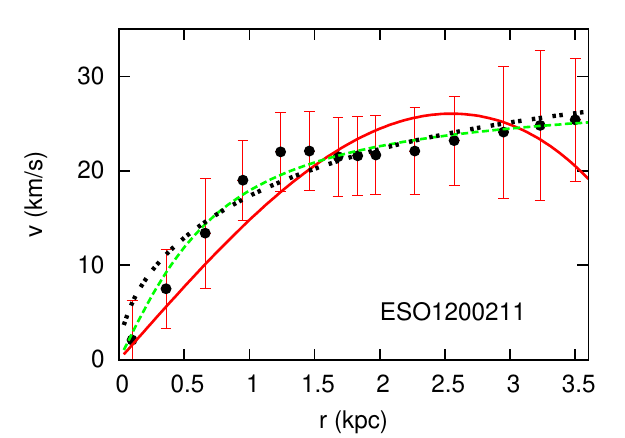}} &
 \resizebox{50mm}{!}{\includegraphics{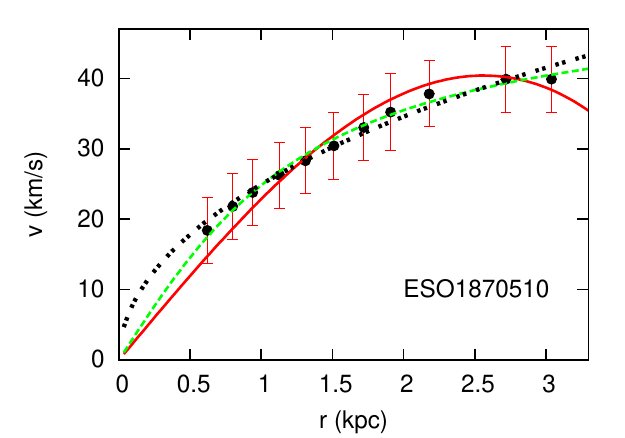}} &
 \resizebox{50mm}{!}{\includegraphics{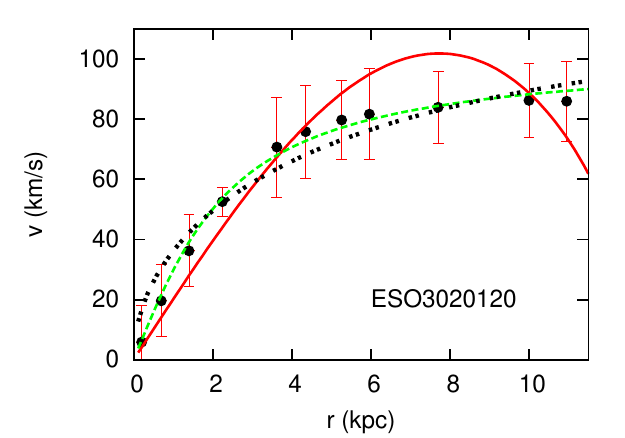}} \\
\resizebox{50mm}{!}{\includegraphics{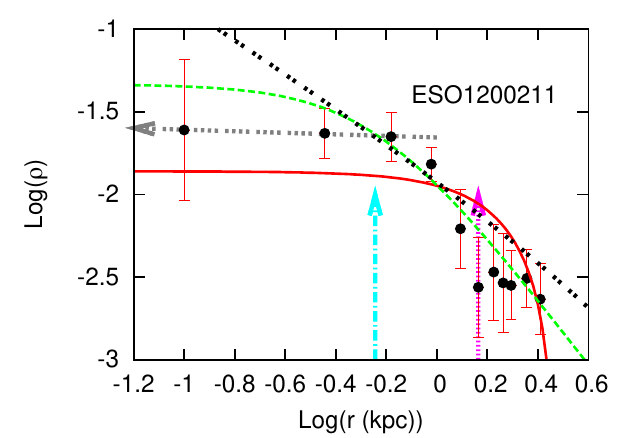}} &
\resizebox{50mm}{!}{\includegraphics{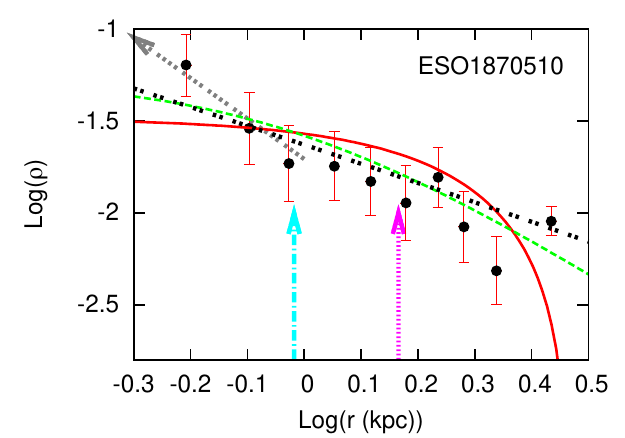}} &
\resizebox{50mm}{!}{\includegraphics{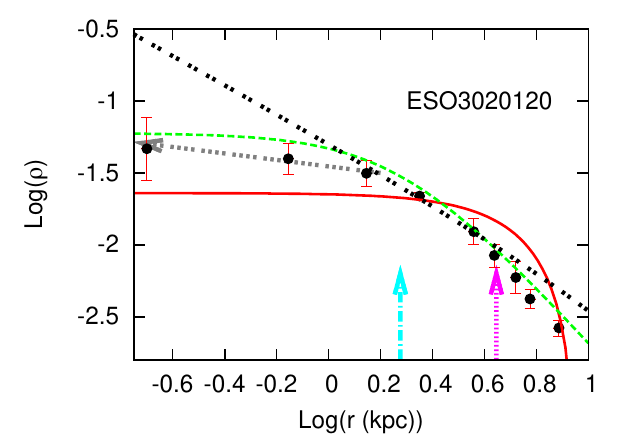}} \\
\resizebox{50mm}{!}{\includegraphics{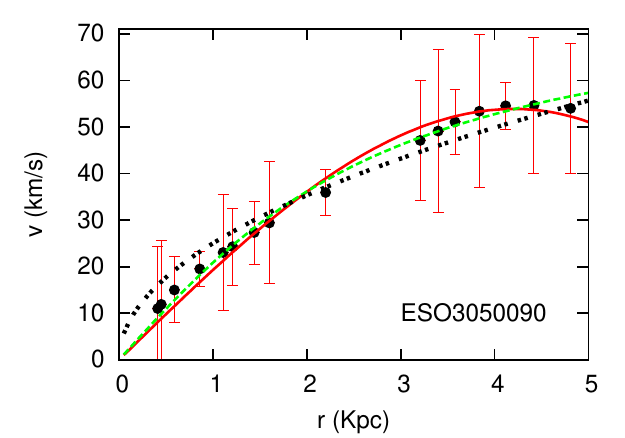}} &
\resizebox{50mm}{!}{\includegraphics{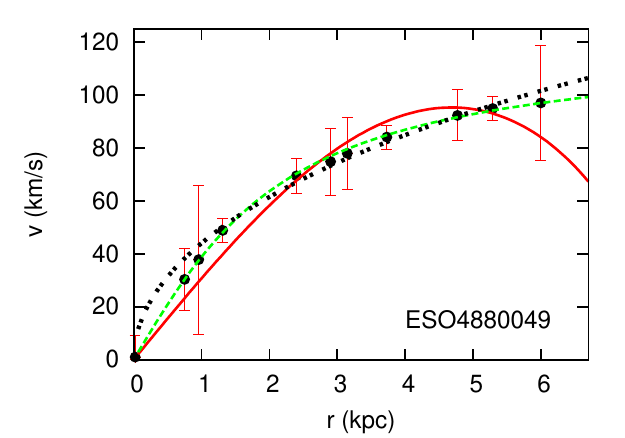}} &
\resizebox{50mm}{!}{\includegraphics{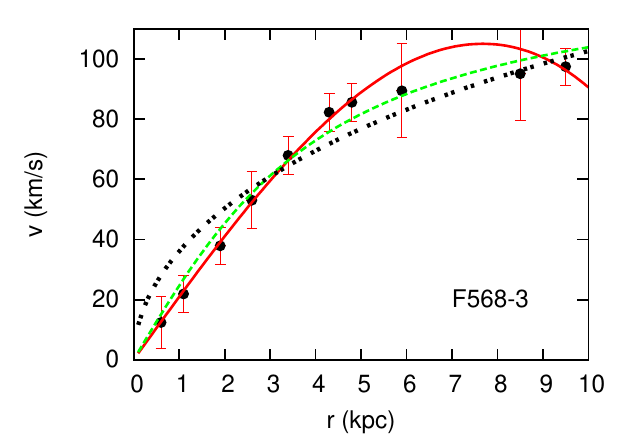}} \\
\resizebox{50mm}{!}{\includegraphics{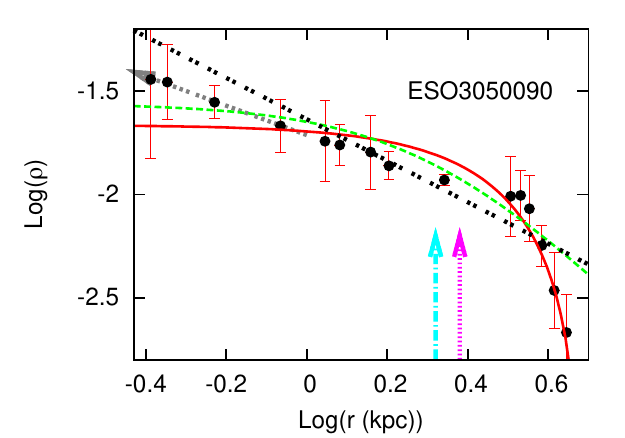}} &
\resizebox{50mm}{!}{\includegraphics{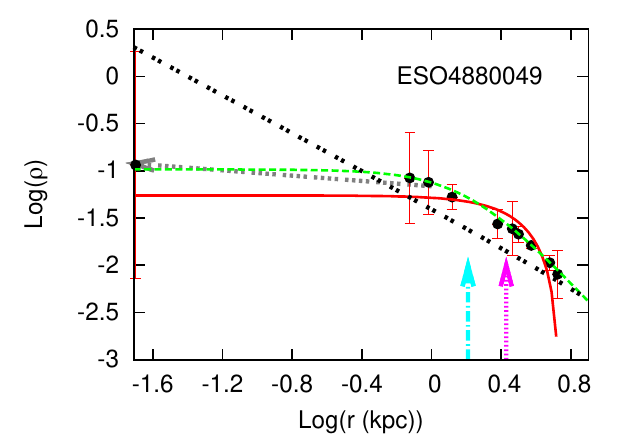}} &
\resizebox{50mm}{!}{\includegraphics{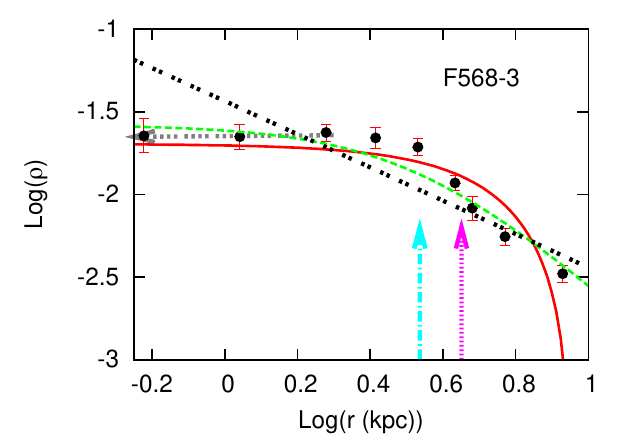}} \\
\resizebox{50mm}{!}{\includegraphics{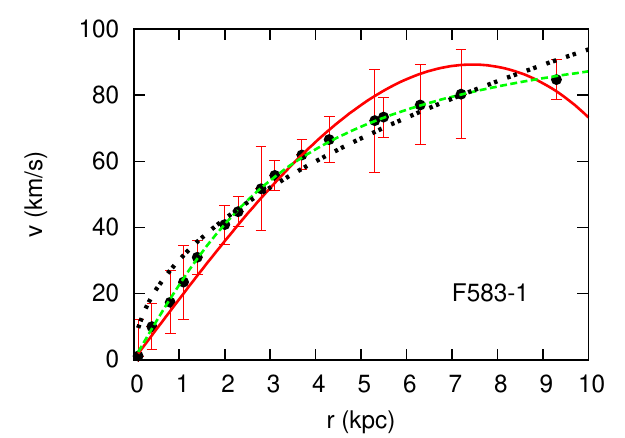}} & 
\resizebox{50mm}{!}{\includegraphics{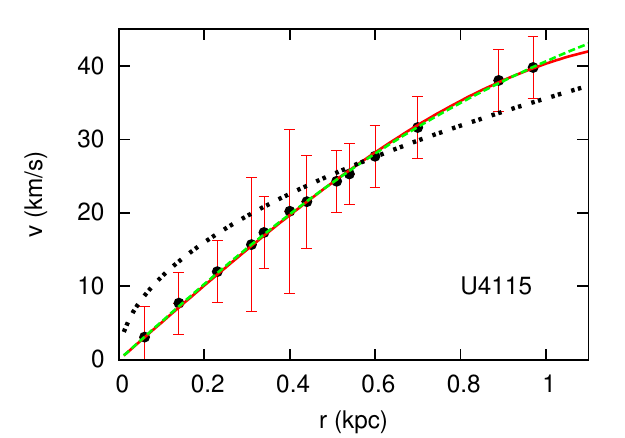}} &
\resizebox{50mm}{!}{\includegraphics{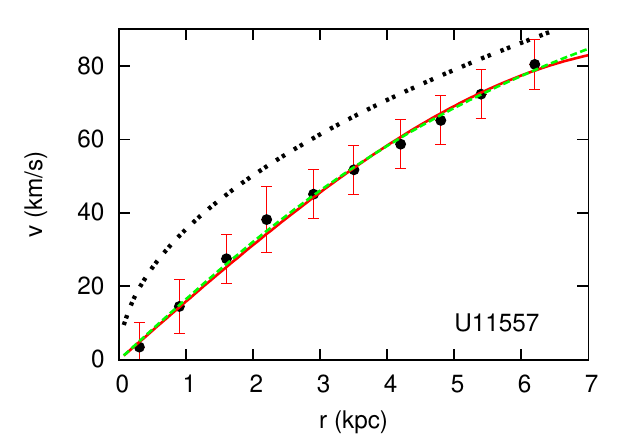}} \\
\resizebox{50mm}{!}{\includegraphics{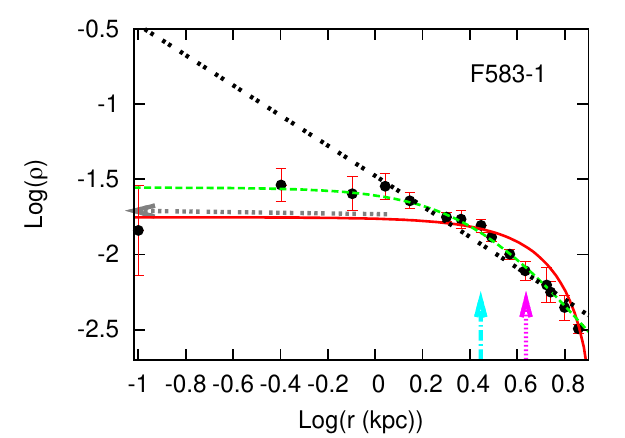}} &
\resizebox{50mm}{!}{\includegraphics{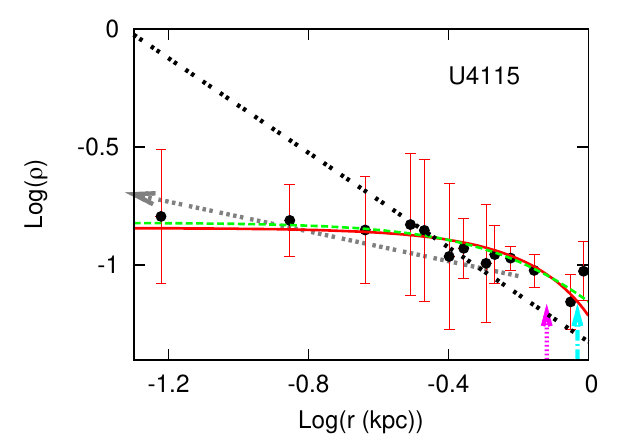}} &
\resizebox{50mm}{!}{\includegraphics{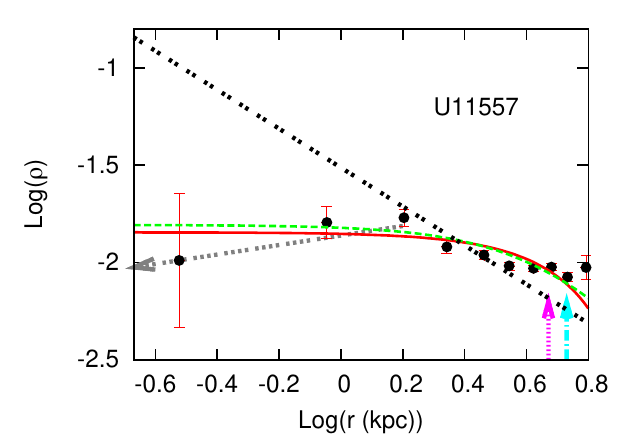}} \\
\end{tabular}
\end{minipage}
\end{figure*}

\begin{figure*}
\begin{minipage}{170mm}
\begin{tabular}{@{}lll@{}}
\resizebox{50mm}{!}{\includegraphics{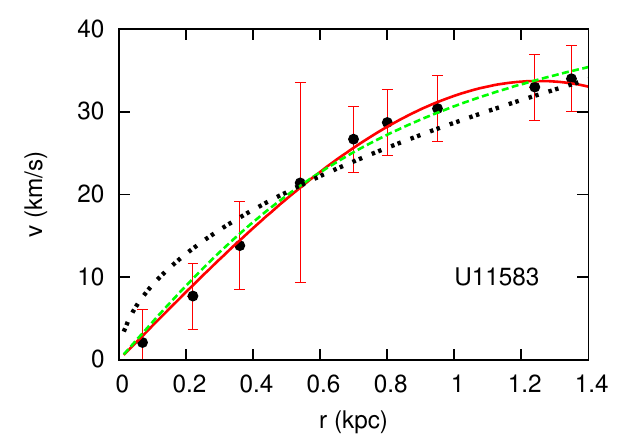}} &
\resizebox{50mm}{!}{\includegraphics{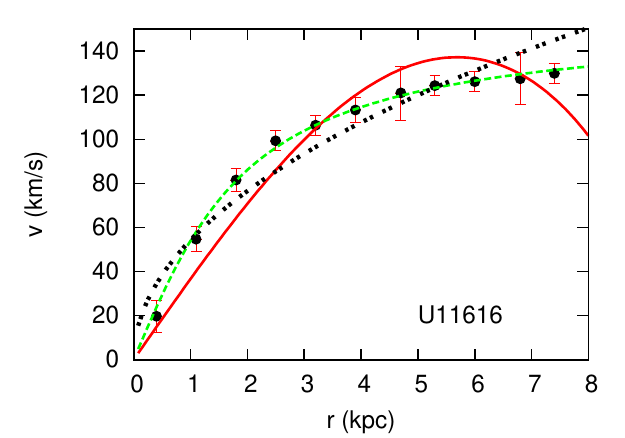}} &
\resizebox{50mm}{!}{\includegraphics{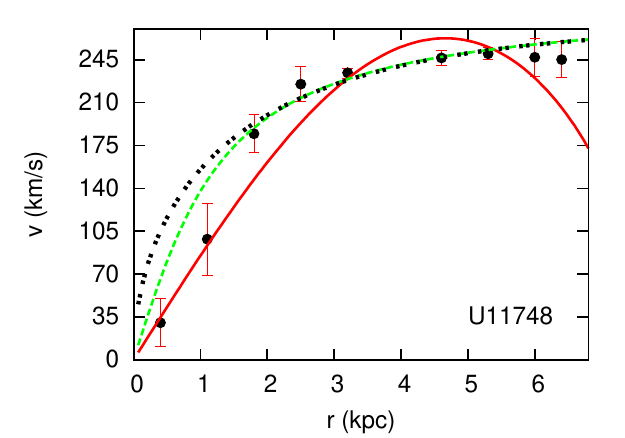}} \\
\resizebox{50mm}{!}{\includegraphics{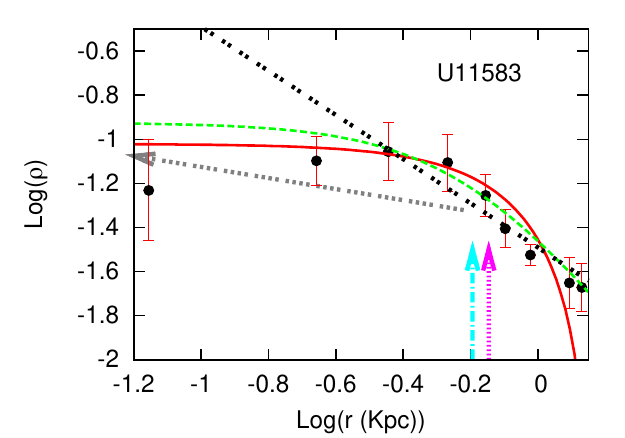}} &
\resizebox{50mm}{!}{\includegraphics{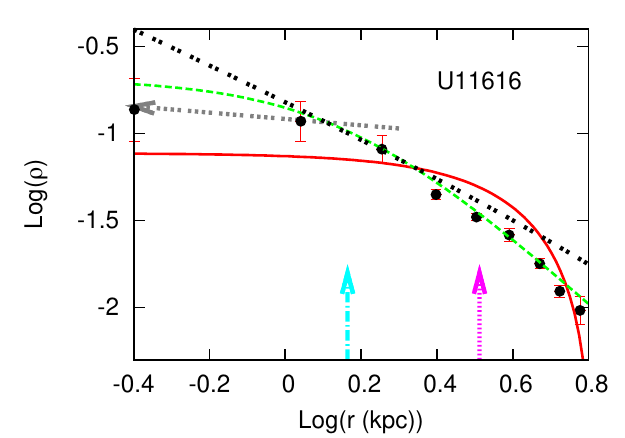}} &
\resizebox{50mm}{!}{\includegraphics{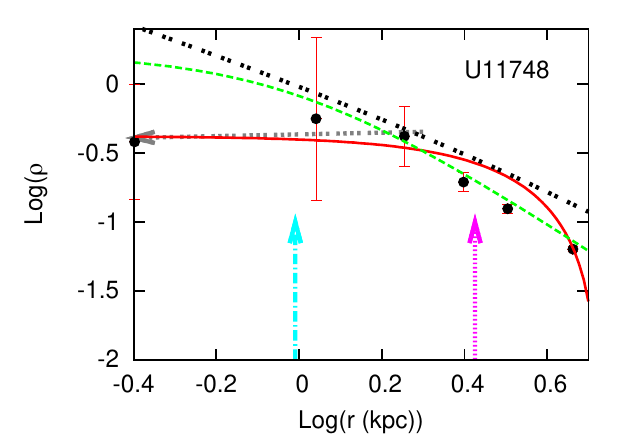}} \\
\resizebox{50mm}{!}{\includegraphics{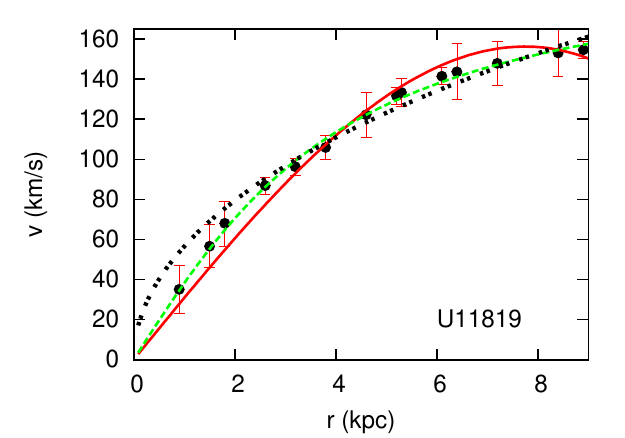}} &
\resizebox{50mm}{!}{\includegraphics{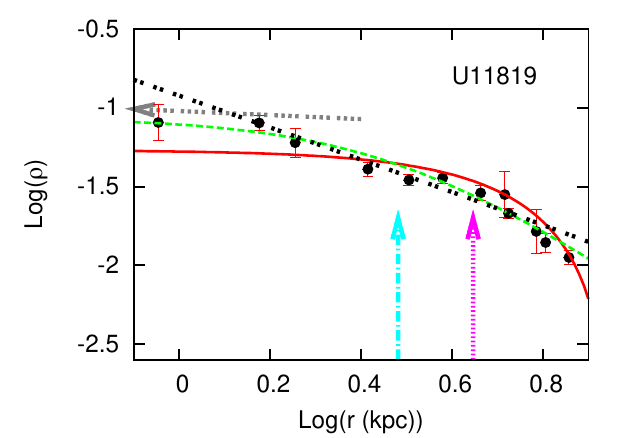}} & \\
\end{tabular}
  \caption{Observed LSB galaxy rotation curves and density profiles with the 
best halo fits. Below each RC is its density profile along with the fits. 
Shown are the PI (green dashed-line); the BEC (red solid line) and 
NFW (black dobble-dotted line) DM halo profiles, the observational data is drawn with error bars. 
The gray arrow denotes the best fit to the data within $R_{1}$ and the vertical arrows 
denote the PI (blue) and BEC (magenta) core radius.}
\label{fig:rotcur}
\end{minipage}

\end{figure*}

\begin{figure}
\centering
 \resizebox{!}{!}{\includegraphics{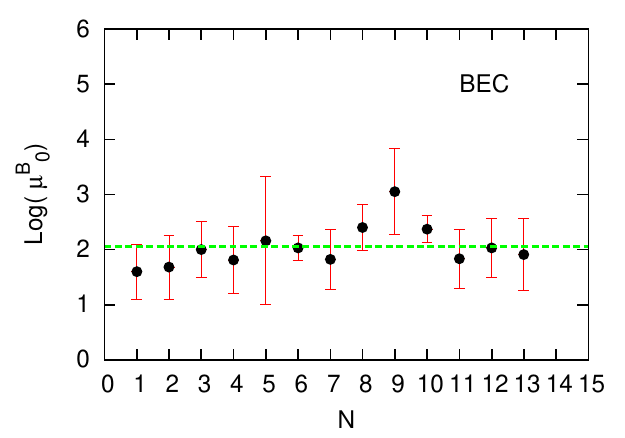}}
 \resizebox{!}{!}{\includegraphics{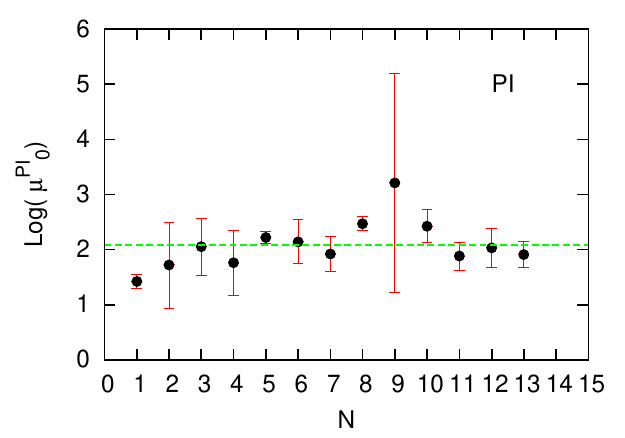}}
  \caption{Plot of log $\big( \mu^{B}_{0}/M_{\odot}pc^{-2} \big)$ and 
 log $\big( \mu^{PI}_{0}/M_{\odot}pc^{-2} \big)$ for each galaxy. N is an identifier for each
galaxy. Here we observe that these values remain 
approximately constant in both profiles, this also serves as a 
crosscheck for our definition of $R_{B}$ in the SFDM/BEC profile.
The green dashed-line represents the mean values given in (\ref{eq:muB}) and (\ref{eq:muPI}).}
\label{fig:logs}
\end{figure}

\subsection[]{Discussion}
The fits of the RCs in Fig \ref{fig:rotcur} prove that the solid-body like behavior characterized by a linear increase in the velocity 
in the central region is more consistent with the core PI and SFDM/BEC profiles than the cuspy NFW. 
The fits within $R_{1}$ give an average value of $\alpha$=$-$0.27 $\pm$ 0.18 consistent 
with those obtained in \citet{deblok2} $\alpha$=$-$0.2 $\pm$ 0.2 and with 
$\alpha$=$-$0.29 $\pm$ 0.07 reported by \citet{ohse} analyzing 7 THINGS dwarf galaxies. 
The case of ESO1870510 might be considered to be consistent with NFW profile, however 
it is the innermost value that considerably decreases $\alpha$, being an irregular galaxy 
more central data near the innermost region is required to discard the possibility of any violent event that 
might have caused such a slope value.

The density profiles corresponding to the RCs fits are also shown in Fig \ref{fig:rotcur} for each galaxy. We see that  
the SFDM/BEC fit slightly deviates from the farthest data points as a result of having a finite 
radius $R$ that is fixed by the same data. This discrepancy is due to the fact that the halo might be more extended
than the value $R$. As a matter of a fact, the more the extended the ``flat'' outer region in the RCs
the more conspicuous the discrepancy. \citet{harkomada} recently found finite temperature 
corrections to (\ref{eq:PI}), this suffice to alleviate the latter problem in LSB galaxies and dwarfs but not for bigger galaxies.
Some solutions have been proposed including vortex lattices \citep{rindler,zinner} and adding more nodes \citep{sin,ji_sin}
in the solution of system (\ref{eq:BEC1}) and (\ref{eq:BEC2})), but so far no final conclusion has been reached.

When comparing the SFDM/BEC and PI core radius we find a general difference of $\sim2$kpc, the core size 
in the PI is approximately $2$ kpc smaller than the BEC core size, but the PI central density is larger.
In U4115, U11557 and U11583 both profiles are very similar which results in a similar core and 
central density values, this can also be taken as a consistency check for our core definition. 

Comparing the values of $R_{B}$ we did not find a tendency to a common value. 
Assuming that the core radius determines the transition where the DM distribution changes from the outer 
region to the inner constant central density, the lack of the unique value means that there is not 
a common radius at which this transition takes place.  

For the second test \cite{victor} uses $R_{B}$ to calculate (\ref{eq:mu}). We have already seen that $R_{c}$ and  
$R_{B}$ are generally different and $R_{B}$ is not a fit parameter. Hence \textsl{a priori} $R_{B}$
is not expected to correlate with $\rho^{B}_{0}$. With the fitting parameters it was obtained
\begin{equation}
\log (\mu^{B}_{0}/ M_{\odot} pc^{-2}) = \log \rho^{B}_{0} R_{B} = 2.05 \pm 0.56, \label{eq:muB}
\end{equation}
\begin{equation}
\log (\mu^{PI}_{0}/ M_{\odot} pc^{-2}) = \log \rho^{PI}_{0} R_{c} = 2.08 \pm 0.46, \label{eq:muPI}
\end{equation}
for the average values in the SFDM/BEC and PI profiles respectively. We see the excellent agreement of 
(\ref{eq:muB}) with (\ref{eq:muPI}) that 
was used as a crosscheck and with (\ref{eq:logmu}) in which a much bigger sample was used.  
The agreement has shown that the SFDM/BEC model is capable of reproducing the constancy 
of the value $\mu_{0}$, something that because of the cuspy nature is not possible in the NFW profile.

In Fig. \ref{fig:logs} we plot the above values for each galaxy. If we define the DM central surface density mentioned
in the introduction, we have for the BEC profile 
\begin{equation}
<\Sigma>^{B}_{0,DM} = M_{< R_{B}} / \pi r^{2}_{B},
\end{equation}
where $M_{< R_{B}}$ is obtained from (\ref{eq:BECm}) evaluated at $R_{B}$.
We show that for U11748 the value $ log \mu^{B}_{0}$ is considerably above the rest and 
with the largest uncertainty. For this reason, in the following analysis we omit both, this value and 
the smallest one that corresponds to ESO1200211. Doing this we get $<\Sigma>^{B}_{0,DM} \approx 191.35$ M$_{\odot}/$pc$^{2}$, 
and for the acceleration felt by a test particle located in $R_{B}$ due to DM only we 
have $g_{DM}(R_{B}) \approx $ 5.2 $\times 10^{-9}$ cms$^{-2}$ broadly consistent with (\ref{eq:gDM}).

The fact that all galaxies present approximately the same order of magnitude in $g_{DM}(R_{B})$ 
might suggest that $R_{B}$ represents more than a transition towards a constant density, it can give us 
information about the close relation between DM and the baryons. Moreover, in view of the lack 
of a unique core radius, we can interpret the transition in DM distribution as an effect 
of crossing a certain acceleration scale instead of a radial length scale. Such interpretation 
reminds us that given in MOND but with the big difference that the acceleration scale found is 
for DM and is not a postulate of the model. 

To determine which interpretation causes the transition whether an acceleration scale or a length scale
we will need to study the properties of larger samples of galaxies with the new telescopes.

\section{Conclusions} \label{sec:conclusions}
In this work, we revisit an alternative paradigm to dark matter nature known as
scalar field dark matter or Bose-Einstein condensate dark matter model. In this model
a fundamental scalar field plays the role of dark matter in the Universe. The hypothesis
is that this scalar field undergoes a phase transition very early in the Universe 
leading to the formation of Bose-Einstein condensate drops. Therefore, the dark haloes of the galaxies in the Universe
are huge drops of scalar field. To explore if this hypothesis is a viable alternative to the standard model we developed
both cosmological and galactic studies.  

In the cosmological regime, we have shown, with a field approach, that the SFDM/BEC model with an ultralight mass of $10^{-22}$ eV 
mimics the behavior of the cosmological expansion rate predicted with the $\Lambda$CDM model. Moreover, although in general a scalar field
is not a fluid, it can be treated as if it behaved like one. Thus, we performed a fluid approach for
the cosmological evolution of the scalar field and we found that the analytical expressions for the kinetic and
potential energies of the scalar field is in excellent agreement with our previous numerical results. 

The interesting cosmological behavior of the scalar field indicates that their scalar fluctuations can be the appropriate 
for the purpose of structure formation, because overdense regions of SFDM/BEC can support the formation of galactic structure.
Thus, we have revisited the growth of SFDM/BEC fluctuations in the linear regime.
Within the linear theory of scalar perturbations we obtain an equation for the evolution of the density contrast. 
This equation differs from the density contrast equation for CDM, however the extra terms tend to the values of the standard equation. 
Therefore, as a goal of this model, we obtained that the scalar perturbations grow up exactly as the CDM paradigm. 
In addition, we study the liner growth of the scalar fluctuations with a fluid approach. In the case
of the SFDM/BEC with $\lambda=0$ we have shown that for the matter dominated era and for big structure this model simulates
the behavior of CDM because in general in a matter dominated Universe for low-$k$, $v_{q}$ tends to 
be a very small quantity tending to zero, so from (\ref{delta}) we can see that on this era we will have the CDM profile 
given by (\ref{deltaCDM}), i.e., the SFDM density contrast profile is very similar to that of the $\Lambda$CDM model (see Fig. \ref{fig2}). 
On the contrary for $\lambda\neq 0$ both models have different behavior as we can see from Fig. \ref{fig3}. 
The numerical results suggest that linear fluctuations on the SFDM can grow faster than those for CDM around $a\sim 10^{-2}$.
Here an important point is that although CDM can grow it does so in a hierarchical way, 
while from Fig. \ref{fig3} we can see that SFDM can have bigger fluctuations just before the $\Lambda$CDM model 
does, i.e., it might be that no hierarchical model of structure formation is needed for SFDM and 
is expected that for the non-linear fluctuations the behaviour will be quite the same as soon as the scalar field condensates, 
in a very early epoch when the energy of the Universe was about $\sim$ TeV. 
Thus, the standard and the  SFDM/BEC model can be contrasted in their predictions concerning the formation of the first galaxies. 
If in the future we see more and more well formed and massive galaxies at high redshifts, 
this could be also a new indication in favour of the SFDM/BEC paradigm. 

On the other hand, we also studied the implications of a SFDM/BEC model at
galactic scales. We find that the SFDM/BEC model gives a constant density profile that is consistent
with RCs of dark matter dominated galaxies. The profile is as good as one of the most
frequently used empirical core profiles but with the advantage of coming from 
a solid theoretical frame. We fit data within $1$kpc and found a logarithmic slope 
$\alpha = -$ 0.27$\pm$ 0.18 in perfect agreement with a core. It is important to notice that the cusp in the central regions 
is not a prediction that comes from first principles in the CDM model, it is a property that is derived by fitting 
simulations that use only DM. In addition, we review a new definition for the core and core radius 
that allows a definite distinction when a density profile is core or cusp. Using this definition we find the core radius 
in the SFDM/BEC profile to be in most cases over $2$kpc bigger than the core radius in the PI profile. 
As a second result of the core definition, we show the constant value of $\mu_{0}$
which is proportional to the central surface density. This result is one of several conflicts 
that jeopardize the current standard cosmological model.

Finally, with all these intriguing results SFDM/BEC model could be a serious alternative to the dark matter problem in the Universe. 
Although the observational evidence seems to be in favor of some kind of cold dark matter,
if we continue to observe even more galaxies at higher redshifts and if higher resolution observations
of nearby galaxies exhibit a core density profile, this model can be a good alternative to $\Lambda$CDM. 
We expect that the close future observations in galaxies surveys can decide the nature of the dark matter.

\section*{Acknowledgments}
%We thank Victor Robles and Abril Suarez for many helpful discussions. 
This work was partially supported by CONACyT M\'exico under grants CB-2009-01, no. 132400, no. 166212, and I0101/131/07 C-234/07 of 
the Instituto Avanzado de Cosmologia (IAC) collaboration (http://www.iac.edu.mx/).

\end{document}